\documentclass{aastex61}
\usepackage{amsmath,amstext}
\usepackage{natbib}
\usepackage{bm}
\accepted{February 26, 2018}
\submitjournal{ApJ}

\shorttitle{Fast funnel-wall accretion onto a weakly magnetized star}
\shortauthors{Takasao et al.}

\begin{document}

\title{A Three-Dimensional Simulation of a Magnetized Accretion Disk: Fast Funnel Accretion onto a Weakly Magnetized Star}

\correspondingauthor{Shinsuke TAKASAO}
\email{takasao@nagoya-u.jp}

\author{Shinsuke TAKASAO}
\affiliation{Department of Physics, Nagoya University, Nagoya, Aichi, 464-8602, Japan}

\author{Kengo TOMIDA}
\affiliation{Department of Earth and Space Science, Osaka University, Toyonaka, Osaka, 560-0043, Japan}

\author{Kazunari Iwasaki}
\affiliation{Department of Earth and Space Science, Osaka University, Toyonaka, Osaka, 560-0043, Japan}

\author{Takeru K. SUZUKI}
\affiliation{School of Arts \& Sciences, University of Tokyo, 3-8-1, Komaba, Meguro, Tokyo, 153-8902, Japan}
\affiliation{Department of Physics, Nagoya University, Nagoya, Aichi, 464-8602, Japan}



\begin{abstract}
We present the results of a global, three-dimensional magnetohydrodynamics simulation of an accretion disk with a rotating, weakly magnetized central star. The disk is threaded by a weak, large-scale poloidal magnetic field, and the central star has no strong stellar magnetosphere initially.
Our simulation investigates the structure of the accretion flows from a turbulent accretion disk onto the star. 
The simulation reveals that fast accretion onto the star at high latitudes occurs even without a stellar magnetosphere.
We find that the failed disk wind becomes the fast, high-latitude accretion as a result of angular momentum exchange mediated by magnetic fields well above the disk, where the Lorentz force that decelerates the rotational motion of gas can be comparable to the centrifugal force. 
Unlike the classical magnetospheric accretion scenario, fast accretion streams are not guided by magnetic fields of the stellar magnetosphere. Nevertheless, the accretion velocity reaches the free-fall velocity at the stellar surface due to the efficient angular momentum loss at a distant place from the star. 
This study provides a possible explanation why Herbig Ae/Be stars whose magnetic fields are generally not strong enough to form magnetospheres also show indications of fast accretion. A magnetically driven jet is not formed from the disk in our model. The differential rotation cannot generate sufficiently strong magnetic fields for the jet acceleration because the Parker instability interrupts the field amplification.
\end{abstract}




\section{Introduction}
Accretion disks are common and fundamental objects in various kinds of astrophysical systems. Circumstellar or protoplanetary disks are formed as a natural consequence of star formation \citep{shu1987,mckee2007}. The disks play roles in feeding central stars \citep[e.g.][]{shakura1973,lynden-bell1974,tomida2017}, controlling the stellar angular momentum \citep{shu1994,hirose1997,matt2005}, and forming planets \citep{hayashi1981,rice2005,machida2011,tsukamoto2015}. Those disks are considered as a driver of jets and outflows \citep{uchida1985,tomisaka2002,fendt2002,pudritz2007}. Accretion processes and acceleration mechanisms of the outflows and jets have also been extensively discussed for accretion disks around compact objects such as black holes and neutron stars \citep{blandford1982,hawley2015}. 

It is well known that an accretion process strongly depends on a magnetic field. If a sufficiently ionized disk is threaded by a weak magnetic field, the disk will be turbulent through the magnetorotational instability \citep[MRI; ][]{velikhov1959,chandrasekhar1961,balbus1991}. The MRI-driven turbulence is believed to be the major mechanism for angular momentum exchange for driving accretion in various kinds of astrophysical disks. Also, a large-scale magnetic field can remove the energy and angular momentum from the disk in the form of jets, outflows, and winds \citep{blandford1982, pudritz1986, shibata1986}, which can result in accretion in the disk. It has been discussed that the inner disk will be truncated and warped when the disk interacts with a strongly magnetized central object \citep[e.g. classical T Tauri stars (CTTSs), neutron stars; ][]{ghosh1979,koenigl1991}.

A classical picture of accretion onto protostars and pre-main-sequence stars is as follows. In the early stage of the accretion phase, the disk extends to the central star and accretes materials to the star in a quiet manner \citep[so-called boundary layer accretion;][]{popham1993,kley1996}. However, as the star evolves, the star may develop a magnetosphere at some point, possibly as a result of the stellar dynamo. As a consequence, the accretion streams are guided by the magnetic field lines of the magnetosphere \citep{ghosh1979,uchida1984,camenzind1990,koenigl1991}. This ``magnetospheric accretion," originally proposed for accreting neutron stars \citep{ghosh1979}, is believed to be a common accretion type for T Tauri stars \citep{bouvier2007,hartmann2016}. The magnetospheric accretion model predicts nearly free-fall accretion to high latitudes (near the magnetic poles).
The regulation of the stellar angular momentum is tightly related to the star-disk interaction via magnetic fields \citep[e.g.][]{shu1994,matt2005}. Quasi-periodic accretion bursts seen in neutron stars and white dwarfs (WDs) may also be controlled by the magnetospheres \citep{spruit1993,scaringi2017}.

Magnetospheric accretion has been widely considered as the most promising accretion model for CTTSs. 
\citet{hartmann1994} argued that the commonly observed inverse P Cygni profile with redshifted absorptions reaching several hundred kilometers per second (close to the escape velocity) is evidence of fast-falling materials. \citet{calvet1998} theoretically investigated the emission from the accretion column that impacts on the stellar surface and demonstrated that the modeled spectrum agrees with the observed UV spectrum that is emitted from the shocked region. Strong stellar magnetic fields on the order of kilogauss, required for the magnetospheric accretion, are also observed \citep{johns-krull1999,johns-krull2007}. Magnetohydrodynamic (MHD) simulations on magnetospheric accretion have been performed extensively \citep[e.g.][]{miller1997,romanova2009a,zanni2013}. Association with X-ray flaring activities and jet acceleration has also been discussed \citep{hayashi1996,hirose1997,goodson1999a}.

However, we still lack direct evidence of magnetospheric accretion, and most of the observational supports rely on the existence of fast accretion. Observations of CTTSs have not clearly supported the scaling relations predicted by the magnetospheric accretion scenario \citep{johns-krull2002}. Most of Herbig Ae/Be stars (HAeBes) do not seem to have the strong magnetic fields required to develop the stellar magnetospheres. The fraction of strongly magnetized ($\gtrsim100~$G) HAeBe stars is only 10~\% \citep{wade2007}. Nevertheless, very weakly magnetized HAeBe stars also show fast accretion \citep{cauley2014}. The accretion signatures are not clearly different between strongly magnetized and very weakly magnetized HAeBe stars \citep{reiter2017}. These observations pose a question on the applicability of the magnetospheric accretion scenario, particularly to HAeBe stars.

The accretion structure is important not only for the stellar evolution but also for the disk evolution. Accretion streams or a warped inner disk can screen out the stellar radiation to the outer disk. This screening effect has impact particularly on the accretion structure by changing the ionization degree in cold disks such as protoplanetary disks, where nonideal MHD effects are significant \citep[e.g.][]{sano2000,sano2002,simon2015,bethune2017,bai2017,suriano2017}. The disk dispersal driven by the photoevaporation wind \citep{hollenbach1994,owen2011,alexander2014} could also be influenced by the screening effect. Optical and infrared observations of accreting CTTSs present many fading events caused by occultation of the stars \citep{bouvier1999,cody2014,stauffer2014}.

Since the accretion process is a nonlinear three-dimensional (3D) process in which magnetic fields play important roles, we need 3D MHD simulations for understanding the accretion structure. 
Most of the previous 3D MHD simulations of the star-disk interaction have been performed on the basis of the magnetospheric accretion scenario \citep{romanova2008,romanova2011}. However, it is still unresolved when and how stars develop their own stable magnetospheres during their evolution. When the stellar magnetic field strength is moderate, the accreting materials will gradually penetrate into the stellar magnetosphere in a fragmented form \citep{stehle2001,kulkarni2008,blinova2016}, in which case the accretion speed will be significantly smaller than the escape velocity. Therefore, it remains unclear if the magnetospheric accretion is a unique solution that realizes a fast accretion at high latitudes, particularly for stars with weak magnetic fields.

In this study, we investigate the accretion from an MRI turbulent disk to a central star without a stellar magnetosphere. This situation is rather simpler than the case of a typical magnetospheric accretion and is very relevant to accreting stars with a weak magnetic field like HAeBes. 
This paper is organized as follows. We will briefly describe our numerical setup and the method of data analysis in Section~\ref{sec:setup}. More detailed information will be given in the Appendix. Section~\ref{sec:results} will show our numerical results of the magnetic field and the accretion structures around the star. In Section~\ref{sec:discussion}, we will compare our results with previous studies and discuss implications for protoplanetary disks. Section~\ref{sec:summary} summarizes our findings.

\section{Numerical Setup}\label{sec:setup}

\subsection{Method and Initial and Boundary Conditions}
Here we give a brief description of our numerical method and initial and boundary conditions. A more detailed explanation is given in the Appendix. 

We solve the three-dimensional MHD equations in a conservative form using Athena++ (Stone et al. 2018, in preparation) in order to simulate the accretion from a disk to a rotating central star. We use the second-order piecewise linear reconstruction method, the  Harten-Lax-van Leer Discontinuities (HLLD) approximate Riemann solver \citep{miyoshi2005}, and the constrained transport method \citep{stone2009} to update the MHD equations. We adopt the equation of state for an ideal gas.
We include a simplified radiative cooling term for the disk material in the energy equation in order to prevent a significant increase in the disk temperature due to viscous heating and to obtain a quasi-steady state for the disk with the initial temperature profile. The reader is referred to Appendix~\ref{apsec:radiative-cooling} for a further information about the cooling.

The initial condition is based on a self-similar, axisymmetric hydrostatic gas distribution consisting of a cold disk and a hot outer atmosphere (see Appendix~\ref{apsec:ic} for more details). We give an hourglass-shaped poloidal magnetic field to the initial atmosphere. The temperature, density, and magnetic field strength profiles on the disk midplane are given so that the sound speed and the Alfv\'en speed scale as the Keplerian velocity. We assume this self-similarity to simplify the initial condition as much as possible. The initial plasma $\beta$ on the midplane is constant with radius and set to $10^4$.

The outer boundary condition is an outgoing boundary at which the inward radial velocity is set to zero. The inner boundary condition, which represents the stellar ``surface", needs to be set with great care, because the inner boundary can be easily numerically unstable due to, for example, the emergence of very low plasma $\beta$ regions and the collision of supersonic accretion streams if any. We should therefore construct a numerically stable inner boundary condition that is physically plausible as well. 

Our stellar surface model is constructed to satisfy the following requirements: (1) the stellar surface should be rigid in the sense that the falling material cannot freely penetrate into the stellar interior, (2) the accreting material will be absorbed by the star eventually but gradually, and (3) the (thermally driven) stellar wind blows from the hot stellar corona. Note that a simple reflecting boundary condition is inappropriate because the accreting material continues to accumulate around the star. The accumulation leads to the formation of an expanding, dense stellar atmosphere, which observations do not support. We include the hot stellar corona because it is commonly observed toward young stars: for CTTSs, see, for example, \citet[][]{feigelson1981,preibisch2005}, and for HAeBes, see, for example, \citet{hamaguchi2005}. To meet the above requirements, we define a so-called damping layer as a thin spherical shell around the actual inner boundary. Physical quantities such as the density and velocity are controlled in the damping layer smoothly with space and time. In terms of the kinematics, the damping layer can be regarded as a viscous layer that connects the rotating star with the outer region. The damping layer physically corresponds to the bottom region of the stellar corona. Hereafter, we refer to the surface of the damping layer as the ``stellar surface," although the stellar surface usually means the stellar photosphere. For more detail, see Appendix~\ref{apsec:damping}.

\begin{figure}
\epsscale{1.0}
\plotone{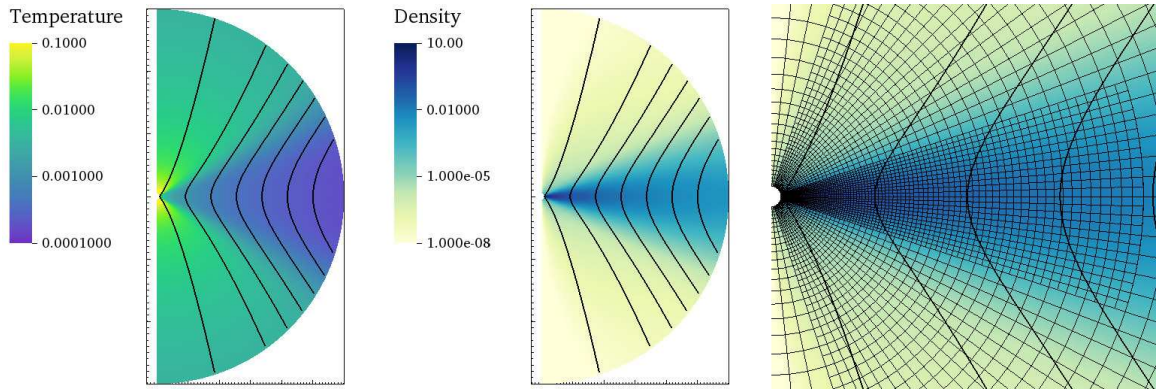}
\caption{Initial setup of the reference simulation. The color contour shows the density. The solid lines represent magnetic fields. The left panel shows the whole simulation domain, while the right panel displays the central part with the mesh structure.}\label{fig:ic}
\end{figure}

The inner boundary condition is as follows. The azimuthal velocity component at the inner boundary is fixed to this stellar rotation velocity. The star is rotating at an angular velocity of $\Omega_{*}$ about the pole. The corotation radius is set to $3R_{*}$, and the resulting $\Omega_{*}$ is $\sqrt{GM_{*}/27R_{*}^3}$, where $G$ is the gravitational constant, $M_*$ is the stellar mass, and $R_{*}$ is the stellar radius. The radial and latitudinal velocity components are set to zero. The magnetic fields in the ghost zones are fixed with time.

We use spherical polar coordinates $(r,\theta,\phi)$ to model the accretion disk around the rotating star. The full solid angle $4\pi$ is covered, which prevents the loss of a poloidal magnetic field from the polar boundaries. The simulation domain is $(0.91R_{*},0,0)\le (r,\theta,\phi) \le (60 R_{*},\pi,2\pi)$. 
The radial grid size is proportional to the radius to keep the ratio between the radial grid size and the longitudinal grid size, and $dr_{\rm i+1}/dr_{\rm i}\simeq1.007$.
We use static mesh refinement to capture MRI-driven turbulence in the disk. The simulation domain is resolved with $120\times 120\times 64$ grid cells at the root level (level~0). The refinement level increases toward the inner disk midplane, and the maximum level is 2. The typical resolution near the midplane is approximately $0.014 R_{*}$ at $r=R_{*}$, which corresponds to 10~cells per local pressure scale height. The initial density and temperature distributions and magnetic field structure are shown in the left panel of Figure~\ref{fig:ic}. The right panel displays the mesh structure.

\subsection{Normalization Units}

\begin{deluxetable*}{ccccc}[b!]
\tablecaption{Normalization units and representative physical values \label{tab:normalization}}
\tablewidth{0pt}
\tablehead{
\colhead{Quantity} & \colhead{Unit} & \colhead{CTTSs ($0.5{\rm M_\odot}$)} & \colhead{HAeBes ($3{\rm M_\odot}$)} & \colhead{WDs ($1{\rm M_\odot}$)}
}
\startdata
Length & $L_0 = R_*$ & $1.4\times 10^{11}$~cm & $1.74 \times 10^{11}$~cm & $5.0\times 10^8$~cm\\
Velocity & $v_0 = v_{\rm K0}$ & $2.2\times 10^{7}$~cm~s$^{-1}$ & $4.8\times 10^{7}$~cm~s$^{-1}$ & $5.2\times 10^8$~~cm~s$^{-1}$\\
Time & $t_0 = R_*/v_{\rm K0}$ & $6.4\times 10^3$ ~s & $3.6\times 10^3$~s & 0.96~s\\
Density & $\rho_0$ & $1.7 \times 10^{-10}$~g~cm$^{-3}$ & $1.7 \times 10^{-10}$~g~cm$^{-3}$ & $1.2\times 10^{-9}$~g~cm$^{-3}$\\
Magnetic Field Strength & $B_0 =  \sqrt{4\pi \rho_0}v_{\rm K0}$& $1\times 10^3$~G & $2.2 \times 10^3$~G & $6.2\times 10^4$~G\\
Mass accretion rate & $\dot{M}_0 = \rho_0 L_0^3 / t_0$ & $1.8 \times 10^{-7} ~{\rm M_{\odot}~yr^{-1}}$ & $6.1 \times 10^{-7}~{\rm M_{\odot}~yr^{-1}}$ & $3.8\times 10^{-10}~{\rm M_{\odot}~yr^{-1}}$\\
\enddata
\end{deluxetable*}

The length unit is taken to be the stellar radius $R_*$. The fiducial velocity is the Keplerian velocity on the stellar surface $v_{\rm K0}=(GM_*/R_*)^{1/2}$. These quantities are used to calculate the other quantities such as the fiducial units of time $t_0$ and magnetic field strength $B_0$. Normalization units and representative physical values are summarized in Table~\ref{tab:normalization}. In the following, we often use time normalized by the Keplerian orbital period at $R=R_*$, $t_{\rm K0}=2\pi t_0$, for convenience.

\subsection{Data Analysis and Averaging Quantities}
In our analysis, we sometimes use cylindrical coordinates $(R,\phi,z)$, although our simulation is performed in spherical coordinates $(r,\theta,\phi)$.

For quantitative analysis, we use physical quantities averaged in the azimuthal and temporal directions. We perform an averaging operation following \citet{suzuki2014}. For variables regarding magnetic fields such as magnetic energy, we take the simple average as follows:
\begin{align}
\langle A \rangle_{t,\phi}=\frac{\int_{t_1}^{t_2} \int_{0}^{2\pi} A dt d\phi}{2\pi(t_2-t_1)},
\end{align}
where $A$ is a physical quantity, and we take the temporal average from $t=t_1$ to $t_2$. For variables regarding velocity such as the flow speed, we conduct the density-weighted average in the following manner:
\begin{align}
\frac{\langle \rho A \rangle_{t,\phi}}{\langle \rho \rangle_{t,\phi}} = \frac{\int_{t_1}^{t_2} \int_{0}^{2\pi} \rho A dt d\phi}{\int_{t_1}^{t_2} \int_{0}^{2\pi} \rho dt d\phi}.
\end{align}

We compare physical quantities in the midplane and around the disk surface to investigate the vertical disk structure. We define the disk surface as the height $|z|=2H_{\rm p}$ from the midplane, since previous studies indicate that the vertical disk structure changes significantly at this height \citep[e.g.][]{suzuki2009,bai2013a}. In our study, one pressure scale height corresponds to the angle of approximately $8^\circ$ ($H_{\rm p}/R \approx 0.14$). We note that this thickness of our disk could be too large for actual disks for CTTSs and HAeBes. For example, when an HAeBe has the stellar mass of $3{\rm M_{\odot}}$ and the disk temperature at the $r=5R_{*}$ is $10^4$~K, $H_{\rm p}/R \approx 0.06$ at this radius. We use that thick disk model to numerically resolve MRI.
For the analysis of the radial dependence, we average quantities around the midplane and disk surface.
The average is performed in the $\theta$ domain of $|z| < H_{\rm p}$ for the midplane region and the domain of $H_{\rm p}<|z|<3H_{\rm p}$ for the disk surface region.


\section{Numerical Results}\label{sec:results}
\subsection{Overview}
\begin{figure}
\epsscale{0.9}
\plotone{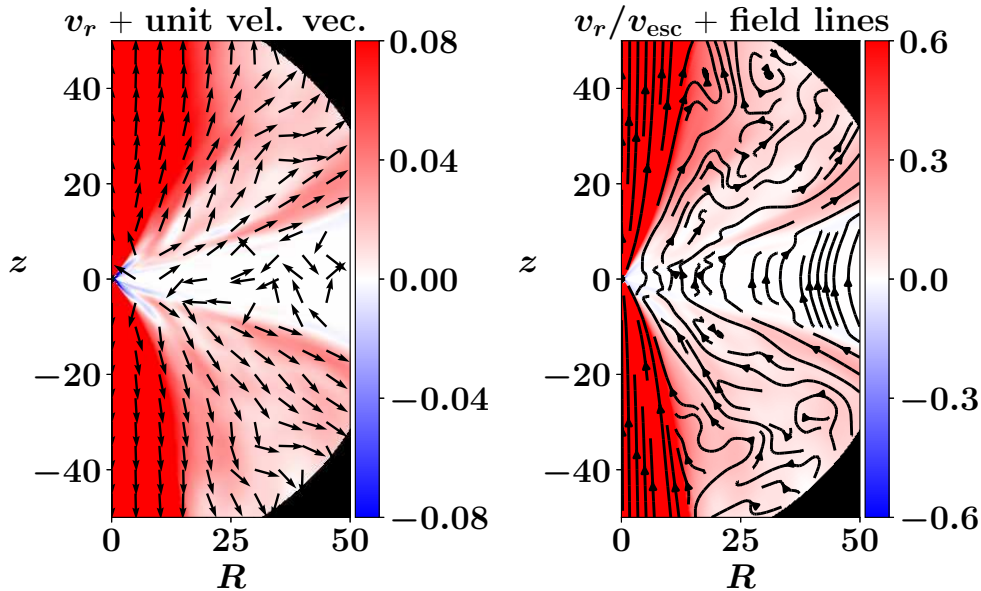}
\caption{Temporally and azimuthally averaged radial velocity (left) and radial velocity normalized by the local escape velocity (right). The arrows in the left panel and the lines with arrows in the right panel indicate the direction of the averaged poloidal velocity (the size does not denote the speed) and averaged poloidal magnetic field lines, respectively. The time average is performed between the times of $t=250t_{\rm K0}$ and $300t_{\rm K0}$.}\label{fig:global-structure}
\end{figure}

We run the simulation until $t=300 t_{\rm K0}$, which corresponds to approximately 27~orbits at $r=5R_*$. A quasi-steady state is achieved within $r\approx 5R_*$.
We first describe the global structures of the magnetic field and the plasma flow. Figure~\ref{fig:global-structure} shows the temporally and azimuthally averaged radial velocity normalized by the local escape velocity. The arrows in the left panel and the lines with arrows in the right panel indicate the direction of the averaged poloidal velocity (their length does not reflect the absolute value) and averaged poloidal magnetic field lines, respectively. The time average is performed between the times of $t=250t_{\rm K0}$ and $300t_{\rm K0}$. A weak wind is driven from the disk. Since the speed is smaller than the escape velocity, the wind is still gravitationally bound in the calculation domain. Strong outflows around the two poles are the thermally driven stellar wind. Magnetic fields above the disk are highly fluctuating, while magnetic fields around the poles are more coherent. 

\begin{figure}
\epsscale{1.2}
\plotone{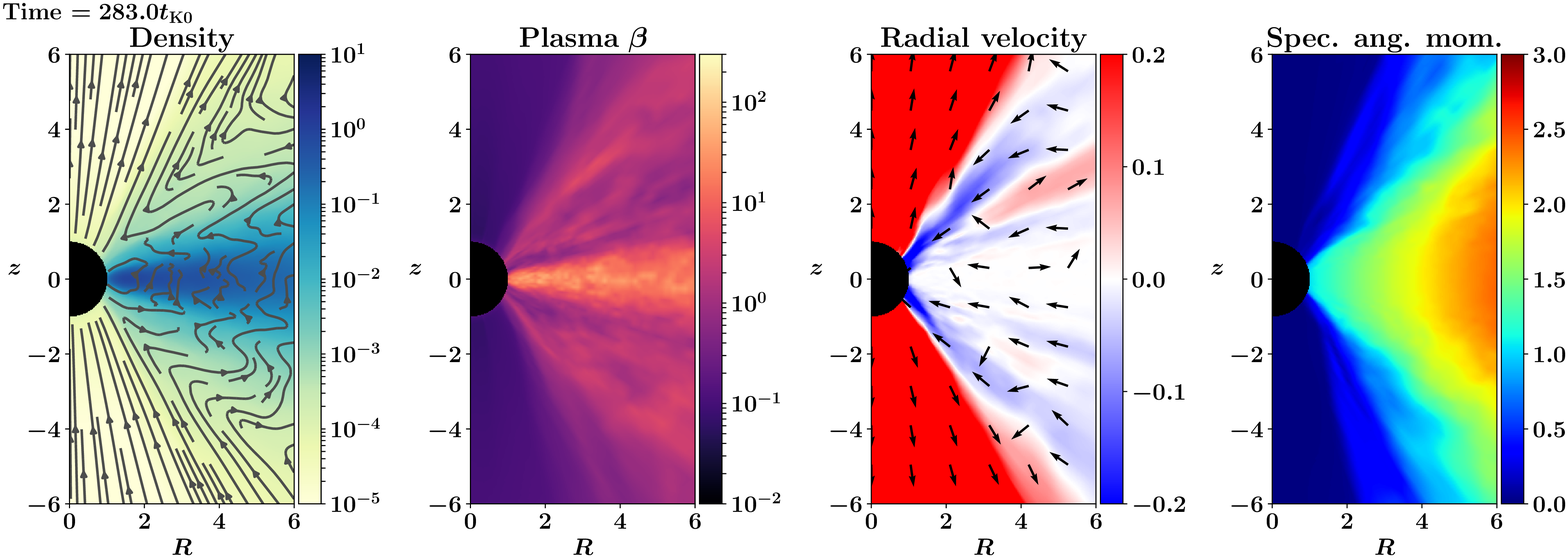}
\caption{From left to right, snapshots of the density, the plasma $\beta$, the radial velocity, and the specific angular momentum at $t=283t_{\rm K0}$ near the star. {The lines with arrows in the density map and the arrows in the radial velocity map indicate the averaged poloidal magnetic field lines and the direction of the averaged poloidal velocity (the size does not denote the speed), respectively.} All of the quantities are averaged in the azimuthal direction. Note that the polar outflows correspond to the thermally driven stellar winds. Animations of this figure are available. The sequence starts at time zero and ends at time 300~$t_{\rm K0}$. The animation duration is 24~s.}\label{fig:overview}
\end{figure}

Figure~\ref{fig:overview} displays the snapshots of the azimuthally averaged density, the plasma $\beta$, the radial velocity, and the specific angular momentum at $t=283t_{\rm K0}$ in the region of $(0,-6R_*)\le (R,z) \le (6R_*,6R_*)$. MRI is developing in the disk from the inner part, making the disk turbulent. The plasma $\beta$ map clearly shows that not only the disk but also the disk atmosphere are highly fluctuating. In addition, the plasma $\beta$ is larger than unity even well above the disk. The disk extends to the stellar surface since the central star in this model does not have a stellar magnetosphere that truncates the inner disk.

Another noticeable feature in Figure~\ref{fig:overview} is the fast accretion flows to the high-latitude regions of the central star (see the radial velocity map). The speed is close to the Keplerian velocity at the stellar surface. The fast, high-latitude accretion is established even without a stellar magnetosphere. On the basis of morphological similarity to funnel-wall jets seen in simulations of black hole accretion disks \citep[e.g.][]{hawley2002}, we call it the funnel-wall accretion.
Vigorous accretion flows around the disk surface are observed, but the funnel-wall accretion flows are distinct from the disk surface accretion flows. The disk surface accretion speed is much smaller than the Keplerian velocity. 
The specific angular momentum map exhibits that the materials of the funnel-wall accretion flows have much smaller angular momenta than those in the disk at the same cylindrical radius. 
In our normalization units, the plasma with the specific angular momentum smaller than unity can fall onto the stellar surface (a detailed analysis will be given in Section~\ref{sec:ang-mom-exchange}).
The funnel-wall accretion is highly time-variable but persistent. We will investigate this accretion in more detail later. 
The spatial resolution diagnostic for MRI is demonstrated in Appendix~\ref{apsec:resolution}. Although the spatial resolution may not be sufficient for the convergence of MRI in the very inner disk, the key physics associated with the funnel-wall accretion are well captured.

\begin{figure}
\epsscale{.50}
\plotone{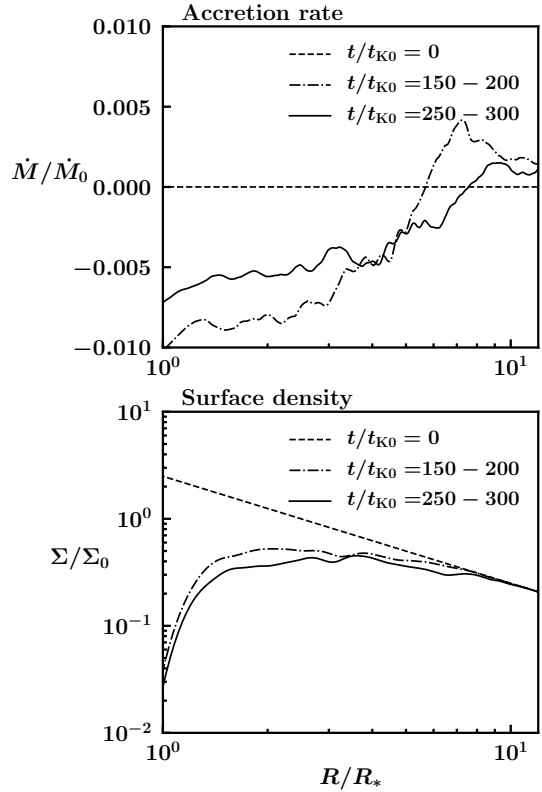}
\caption{Time evolution of the mass accretion rate (top) and the surface density (bottom) as a function of radius. Different lines show data at different timings. The spans for time averaging are shown in the legends. The unit of time is $t_{\rm K0}$, and $\Sigma_0$ is the unit of the surface density and is given by $\rho_0 L_0$.}\label{fig:sigma-mdot}
\end{figure}

The accretion rate as a function of radius is shown in the top panel of Figure~\ref{fig:sigma-mdot}. 
The plot suggests that the inner part accretes and the outer part moves outward, and that the location dividing the inward and outward flows moves outward, which seem to follow the time evolution of the self-semilar solution of the standard accretion disk \citep{lynden-bell1974}. However, we cautiously note that in our simulation the disk mass is partially lost via outflows and  turbulence is still developing in the disk. These effects are not considered in the self-similar solution, and therefore we cannot directly compare the top panel of Figure~\ref{fig:sigma-mdot} to it in a quantitative sense. 
The surface density profile becomes flat, which is similar to the result of the previous study by \citet{suzuki2014}. Since the accretion rate is almost constant within $r\approx 5R_*$, the inner part within this radius reaches a quasi-steady state.

%

\subsection{Vertical Disk Structure}
\begin{figure}
\epsscale{0.8}
\plotone{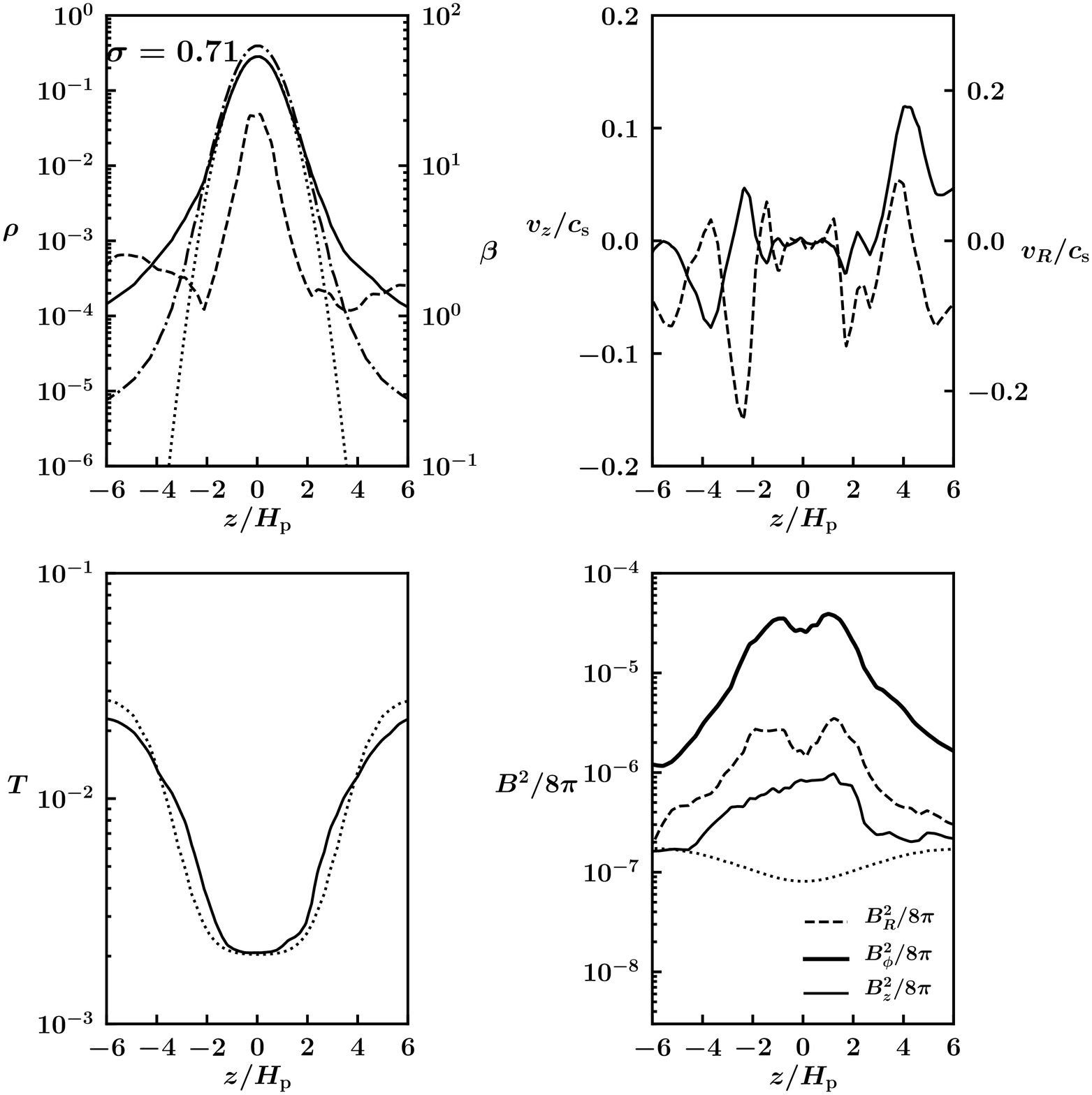}
\caption{Temporally and azimuthally averaged vertical structure of the disk at $R=5R_*$. The horizontal axes show the height from the disk midplane normalized by the pressure scale height $H_{\rm p}$. The time average is performed between $t=250t_{\rm K0}$ and $300t_{\rm K0}$. Top left: the density (solid line, left axis) and the plasma~$\beta$ (dashed line, right axis). The dotted line denotes the Gaussian fitting for the density, which visualizes that the density profile deviates from a Gaussian profile around $z\approx 2H_{\rm p}$. Here, $\sigma$ is the width of the Gaussian profile. The dash-dotted line shows the initial density distribution. Top right: $v_{z}/c_{\rm s}$ (solid line, left axis) and $v_{R}/c_{\rm s}$ (dashed line, right axis). The $v_{z}/c_{\rm s}$ plot shows that the disk wind is subsonic. Signatures of the funnel-wall accretion can be seen around $z\approx \pm 5H_{\rm p}$ as local minima in $|v_{ R}/c_{\rm s}|$, as well as signatures of the disk surface accretion around $z\approx \pm 2H_{\rm p}$. Bottom left: the temperature. The dotted line denotes the initial temperature. Bottom right: the magnetic energy density ($B_{R}^2/8\pi$, dashed line; $B_{ \phi}^2/8\pi$, thick solid line; and $B_{ z}^2/8\pi$, solid line). The dotted line denotes the initial $B_{z}^2/8\pi$.}\label{fig:vertical-distribution}
\end{figure}

We investigate the vertical structure of the disk. Figure~\ref{fig:vertical-distribution} exhibits the vertical profiles of physical quantities. The horizontal axis denotes the height from the disk midplane normalized by the pressure scale height defined at the midplane, $H_{\rm p}=\sqrt{2}c_{\rm s,mid}(R)/\Omega_{\rm K}(R)$, where $c_{\rm s,mid}(R)$ is the sound speed at the midplane and $\Omega_{\rm K}(R)$ is the angular velocity of the Keplerian motion at the cylindrical radius $R$. 

In the top left panel, the density profile (solid line) deviates from a Gaussian profile (dotted) around $z=2H_{\rm p}$. The plasma~$\beta$ falls from the midplane toward that height, but remains within 1 to 10 above the height. The vertical velocity profile in the top right panel indicates that disk materials are lifted from the disk surface, suggesting the mass loading by subsonic upflows, which are driven by the gradient of the turbulent magnetic pressure and magnetic tension (shown later in this section). Although the signature of the upflows is somewhat hidden by the fast accretion within $ R \sim 5R_*$ in the azimuthally and temporally averaged image, the upflows are clearly seen outside that cylindrical radius (see, e.g. Figure~\ref{fig:accretion-structure}).

The bottom left panel of Figure~\ref{fig:vertical-distribution} shows that the temperature in the upper atmosphere does not decrease from the initial value, although cool materials are lifted from the disk. This suggests that the heating by turbulent dissipation takes place in the upper atmosphere \citep{io2014}. The temperature in the disk is maintained at the initial value by the adopted radiative cooling effect.
The bottom right panel shows the three components of the magnetic energy. The toroidal field $B_\phi$ is the most dominant component among the three magnetic field components, as commonly seen in simulations of magnetized disks. The second-largest field component is $B_{R}$, followed by $B_{ z}$. The field amplification mainly occurs at $|z| \lesssim 2H_{\rm p}$.

The plasma $\beta$ is still larger than unity above the disk surface ($z\gtrsim 2H_{\rm p}$) and has a flatter profile there. We note that previous global simulations covering a limited $\theta$ domain showed a much lower plasma $\beta$ near the $\theta$ boundaries \citep{suzuki2014}, although their models assume a weak disk magnetic field (e.g. in \citet{suzuki2014}, $\beta=10^5$). In a case with the polar boundaries near the disk surface, the disk materials that go through the boundaries are extracted from the simulation domain even though they are gravitationally bound. Therefore, the atmosphere near the polar boundaries tends to have an artificially low density and low $\beta$ in such a case. This artificial mass extraction does not occur in our simulation, which covers a solid angle of $4\pi$. We consider that this setting results in a higher plasma $\beta$ in the upper atmosphere.

\begin{figure}
\epsscale{0.8}
\plotone{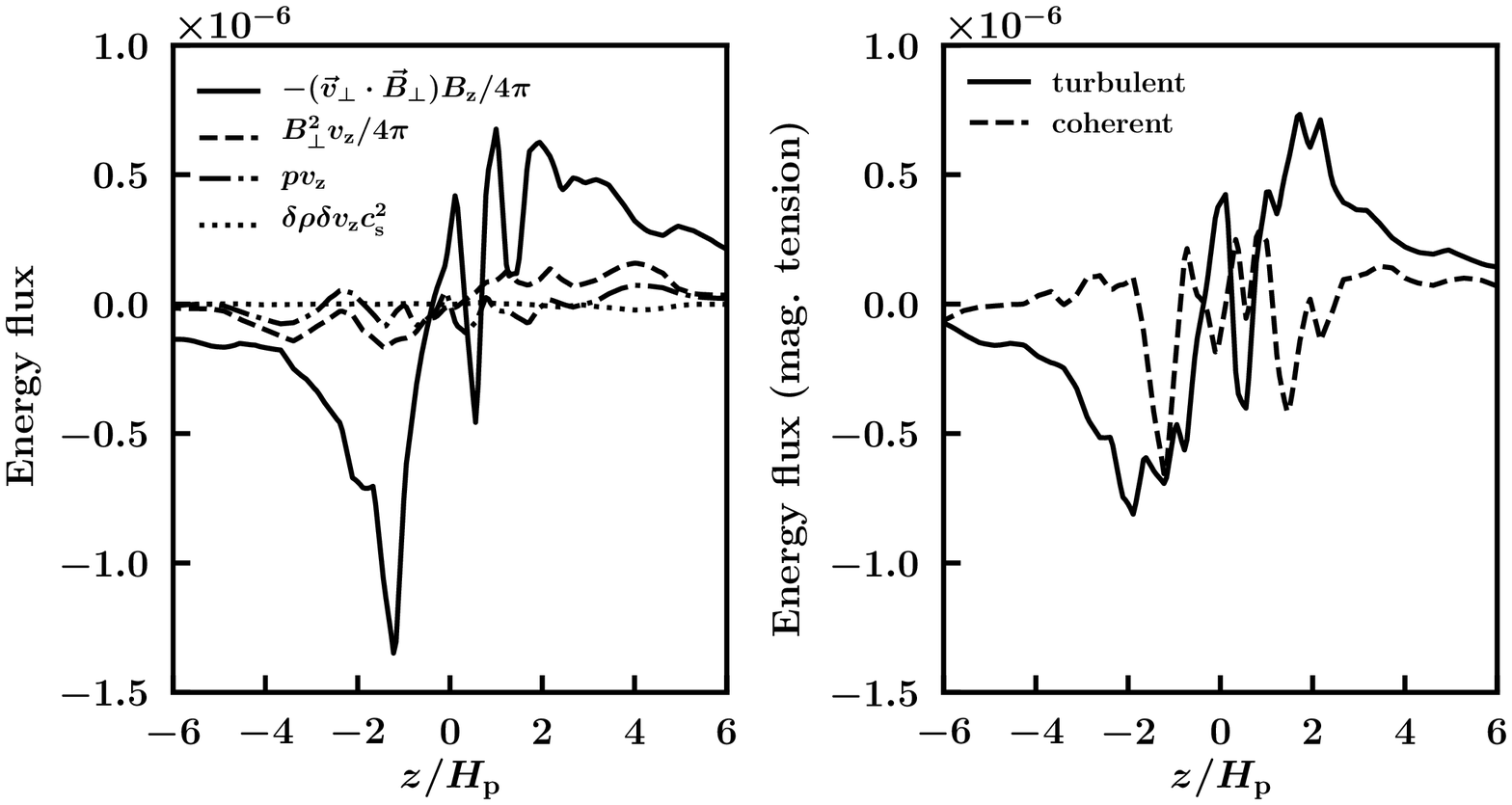}
\caption{Same as Figure~\ref{fig:vertical-distribution}, but for different energy fluxes. Left: vertical energy flux associated with the magnetic tension (solid line), the advection of magnetic energy (dashed), gas pressure (dash-dotted), and sound waves (dotted). Right: turbulent (solid) and coherent (dashed) components of energy flux associated with magnetic tension.}\label{fig:vertical-distribution-Eflux}
\end{figure}

The lifting of the disk material due to MRI turbulence corresponds to the MRI-driven disk wind \citep{suzuki2009,suzuki2014,flock2011,fromang2013,bai2013a}. The launching height ($\sim 2 H_{\rm p}$, see the top right panel in Figure~\ref{fig:vertical-distribution}) is consistent with the previous studies. To investigate the driving process, we analyze energy transfer in the vertical direction shown in Figure~\ref{fig:vertical-distribution-Eflux}. We define $\bm{v}_{\perp} = (v_{R}, v_{ \phi})$ and $\bm{B}_{\perp} = (B_{R},B_{\phi})$ as the velocity and magnetic field vectors perpendicular to the $z$-direction, respectively. We consider the Poynting flux due to magnetic tension ($(\bm{v}_{\perp}\cdot \bm{B}_{\perp})B_{ z}/4\pi$), the Poynting flux due to the advection of the magnetic energy ($B_{\perp}^2 v_{ z}/4\pi$), the power done by the pressure ($p v_{ z}$), and the energy flux of sound waves ($\delta \rho \delta v_{ z} c_{\rm s}^2$). Here, $\delta \rho = \rho - \langle \rho \rangle_{ t,\phi}$ and $\delta v_{ z}=v_{ z}-\langle v_{ z} \rangle_{t,\phi}$, where the time average is performed between $t=250t_{\rm K0}$ and $300t_{\rm K0}$.

The left panel of Figure~\ref{fig:vertical-distribution-Eflux} displays that the magnetic tension term is dominant among them above the disk, which indicates that the weak disk wind is mainly driven by the magnetic tension. The advection of the magnetic energy is much smaller than the magnetic tension, which is different from the result of \citet{suzuki2009} in a quantitative sense. 

The magnetic tension term is decomposed into the turbulent and coherent components in the right panel of Figure~\ref{fig:vertical-distribution-Eflux}. The coherent component is defined as 
\begin{align}
-(\langle \bm{v}_{\perp}\rangle_{t, \phi} \cdot \langle \bm{B}_{\perp}\rangle_{ t, \phi})\langle B_{ z} \rangle_{t, \phi} / 4\pi,
\end{align}
and the turbulent component as 
\begin{align}
-\left[ \langle (\bm{v}_{\perp}\cdot \bm{B}_{\perp})B_{z} \rangle_{t, \phi} / 4\pi - (\langle \bm{v}_{\perp}\rangle_{t, \phi} \cdot \langle \bm{B}_{\perp}\rangle_{t, \phi})\langle B_{z} \rangle_{t, \phi} / 4\pi \right].
\end{align}
The figure indicates that the turbulent component mainly carries the energy to the upper atmosphere. From this fact, the disk wind is driven by the turbulence, and therefore it is the MRI-driven wind. This wind supplies materials to the upper atmosphere.

\begin{figure}
\epsscale{0.6}
\plotone{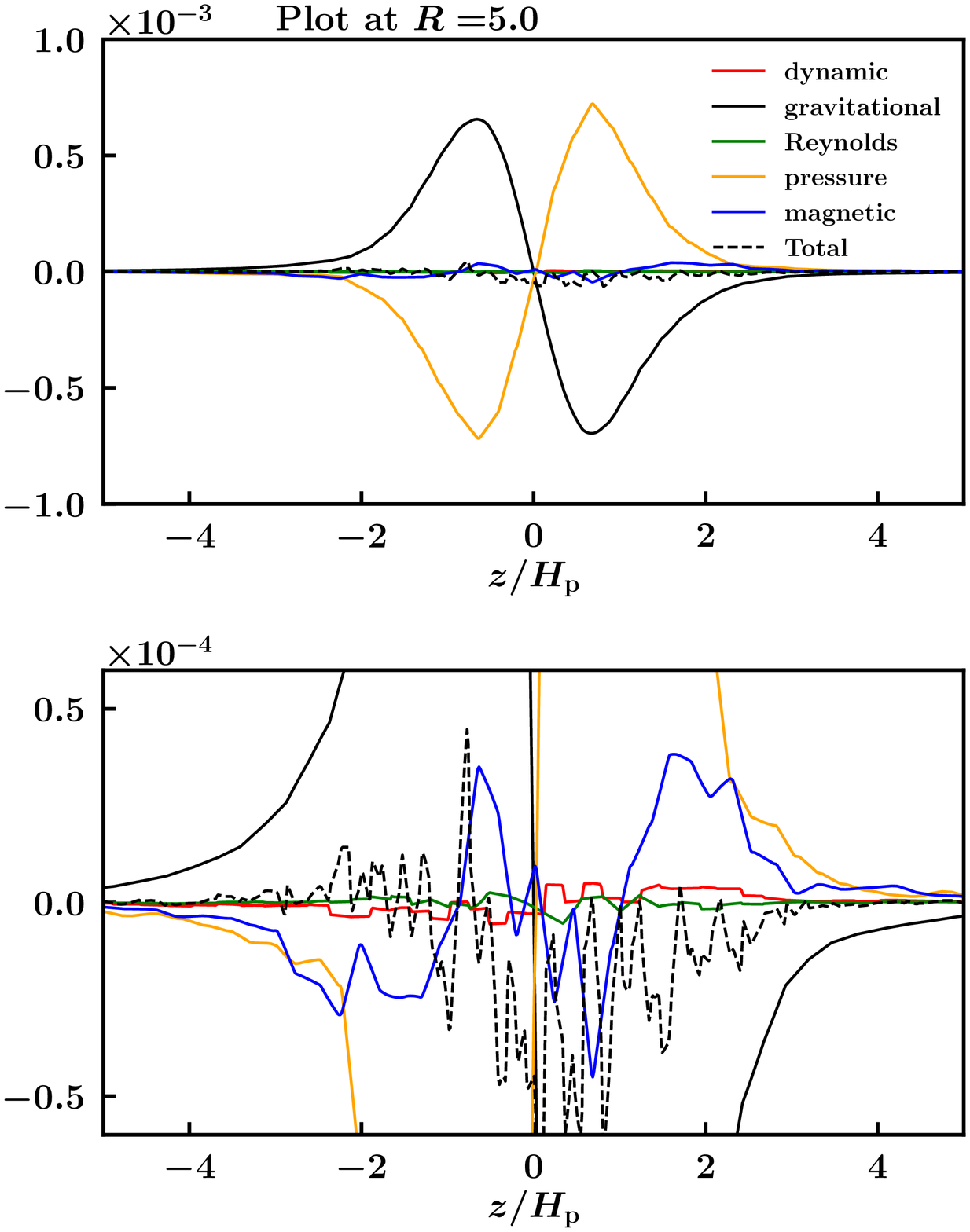}
\caption{Vertical distributions of the different kinds of forces. The gas pressure gradient (orange line), the magnetic pressure gradient associated with the toroidal field $\partial(\langle B_{ \phi}^2\rangle_{ t,\phi}/8\pi)/\partial z$ (blue), the dynamic pressure gradient $\partial (\langle \rho v_{ z}^2\rangle_{t,\phi})/\partial z$ (red), the Reynolds stress $R^{-1} \partial (R \langle \rho v_{ R}v_{ z}\rangle_{ t,\phi})/\partial R$ (green), the gravitational force (solid black), and the total force (dashed black) in the vertical direction at $R=5R_*$ are shown. The bottom panel is the same as the top panel but with a much smaller range of the vertical axis.}\label{fig:vertical-distribution-force}
\end{figure}

The different kinds of forces that determine the vertical stratification are displayed in Figure~\ref{fig:vertical-distribution-force}. The figure shows the gas pressure gradient, the magnetic pressure gradient associated with $B_{ \phi}$, the dynamic pressure gradient $\partial (\langle \rho v_{z}^2\rangle_{ t,\phi})/\partial z$, the Reynolds stress $R^{-1} \partial (R \langle \rho v_{ R}v_{ z}\rangle_{t,\phi})/\partial R$, the gravitational force, and the total force in the vertical direction at $R=5R_*$. Since the total force is much smaller than the gravitational force, the disk and the atmosphere are almost hydrostatic.
As shown in the top panel, the gravitational and the pressure gradient forces are the most dominant terms around the midplane and are almost balanced. However, as expected from Figures~\ref{fig:vertical-distribution} and \ref{fig:vertical-distribution-Eflux}, the magnetic pressure gradient becomes important above the disk surface. The bottom panel is the same as the top panel but with a much smaller range of the vertical axis. The dynamic pressure gradient force and the Reynolds stress in the vertical direction are negligible. The dynamic pressure is much less important than the gas pressure because the MRI-driven wind is subsonic (Figure~\ref{fig:vertical-distribution}).

\begin{figure}
\epsscale{0.6}
\plotone{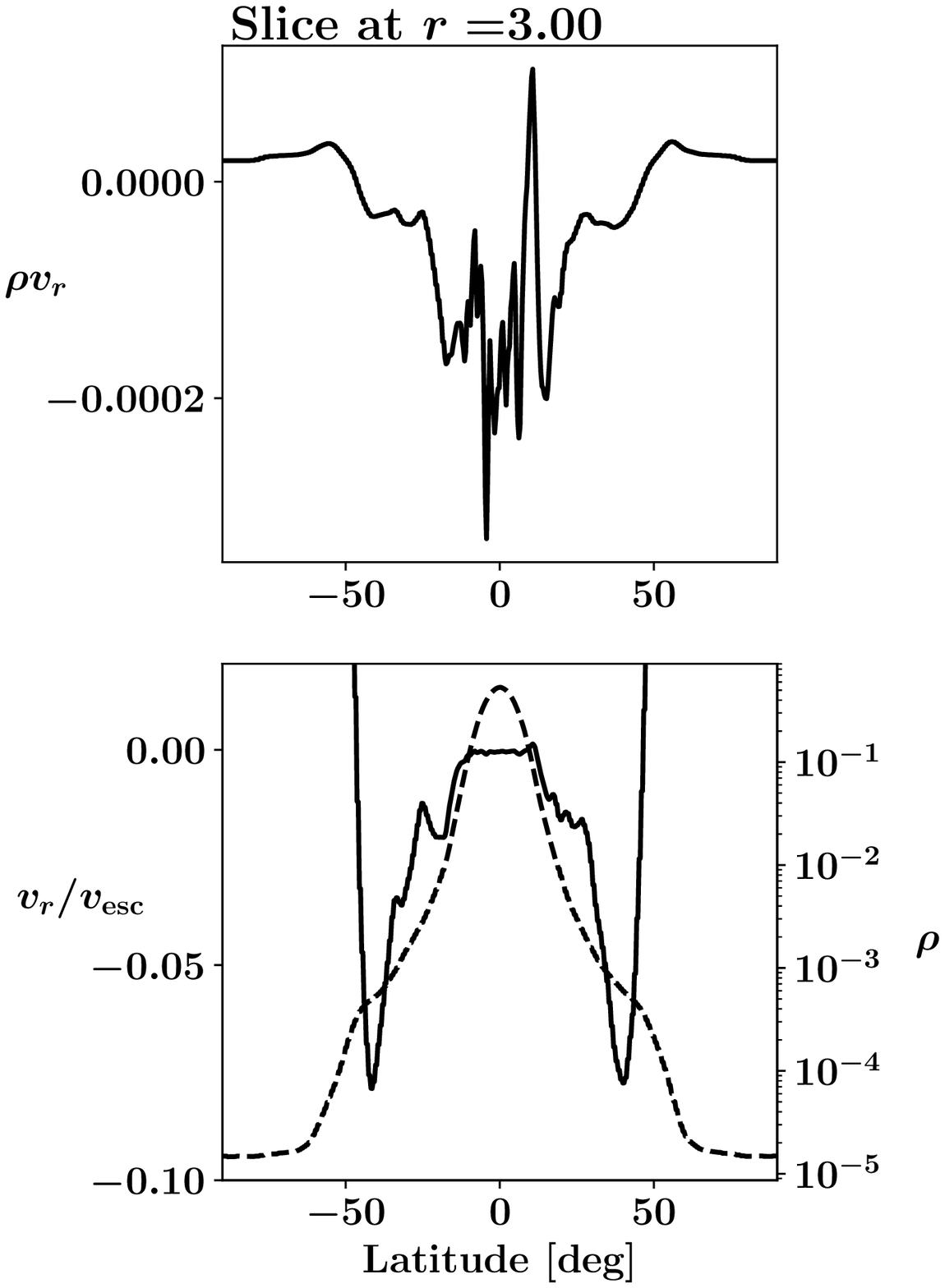}
\caption{Accretion structure. Top: the mass flux. Bottom: the radial velocity normalized by the escape velocity (solid line) and the density (dashed). Note that the ratio of the initial pressure scale height to the radius ($H_{\rm p}/r$) is $\sim 0.14$, which corresponds to $\sim 8^\circ$. The height $z=2H_{\rm p}$ is therefore approximately $\pm16^\circ$ from the midplane. The accretion around those heights is the disk surface accretion. The faster accretion seen at further higher altitudes corresponds to the funnel-wall accretion.
The quantities in this figure are temporally and azimuthally averaged. The time average is performed between $t=250t_{\rm K0}$ and $300t_{\rm K0}$.}\label{fig:accretion-rate-structure}
\end{figure}

The mass accretion actively operates near and above the disk surface. Figure~\ref{fig:accretion-rate-structure} shows the latitudinal distributions of the temporally and azimuthally averaged mass flux, radial velocity, and density, measured at the spherical radius of $3R_*$. One will notice that the incoming mass flux near the disk surface is comparable to that around the midplane. Note that the ratio of the initial pressure scale height to the radius ($H_{\rm p}/r$) is $\sim 0.14$, which corresponds to $\sim 8^\circ$. The height $z=2H_{\rm p}$ is therefore approximately $\pm16^\circ$ from the midplane. A similar surface-layer accretion structure is also seen in previous studies \citep[e.g.][]{suzuki2014}. 
A large accretion speed above the disk (approximately $\pm 40^\circ$) indicates the funnel-wall accretion. While the accretion speed is higher in the funnel-wall accretion, the mass flux (accretion rate) is lower than in the disk and surface accretion because of its low density.

\begin{figure}
\epsscale{0.9}
\plotone{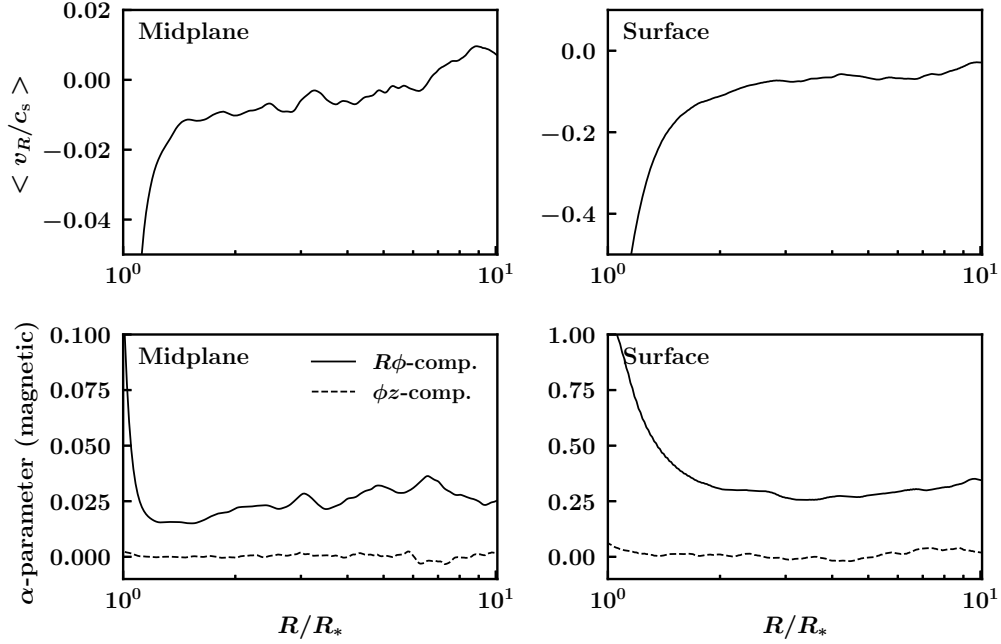}
\caption{Comparison between the disk midplane and the disk surface structures. The top panels show the Mach number of the accretion speeds, and the bottom panels compare the magnetic $\alpha$ parameters (left: midplane, right: disk surface).}\label{fig:accretion-midplane-surface}
\end{figure}

Figure~\ref{fig:accretion-midplane-surface} compares the accretion velocity (top panels) and the so-called $\alpha$ parameter due to the Maxwell stress (bottom panels) around the midplane and disk surface. The $\alpha$ parameters due to the $R\phi$ and $\phi z$ components of the Maxwell stress are written as $\alpha_{R\phi}^{\rm m}=-\langle B_{ R}B_{ \phi}\rangle_{ t,\phi}/4\pi \langle p \rangle_{t,\phi}$ and $\alpha_{ \phi z}^{\rm m}=-\langle B_{\phi}B_{ z}\rangle_{ t,\phi} / 4\pi \langle p \rangle_{ t,\phi}$, respectively.
The accretion speed is much larger around the disk surface than around the midplane. The difference in magnitude of the $\alpha$ parameter between the midplane and the surface accounts for the difference of the accretion speed. The larger $\alpha$ around the disk surface is a result of a flatter profile of the magnetic energy distribution than the density/pressure profiles (see Figure~\ref{fig:vertical-distribution}). 

\begin{figure}
\epsscale{1.10}
\plotone{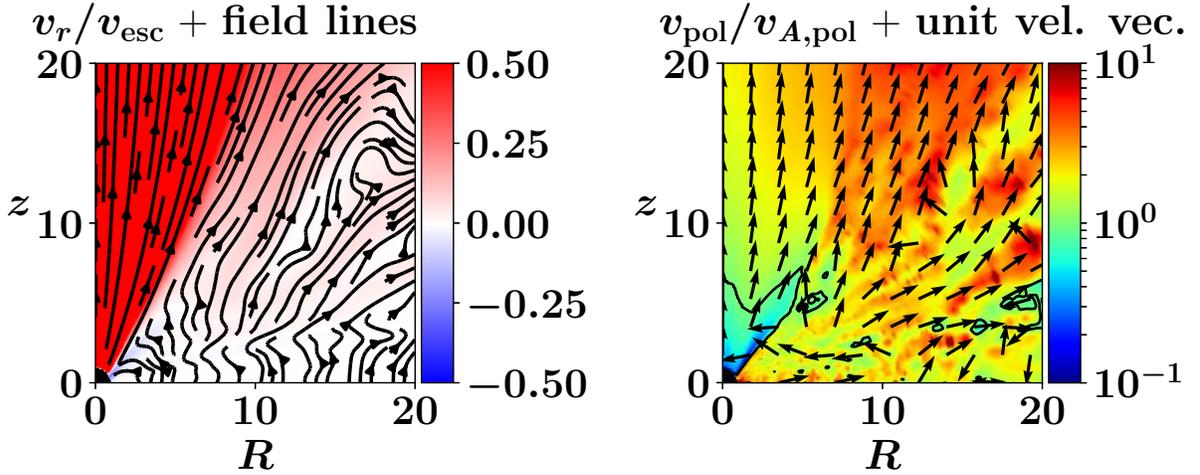}
\caption{Averaged magnetic field and flow structures. Left: contour indicates the radial velocity normalized by the escape velocity. Lines denote magnetic field lines. Right: Alfv\'en Mach number, defined by the ratio of the poloidal velocity $v_{\rm pol}$ to the poloidal Alfv\'en velocity $v_{{\it A}, \rm pol}$, is shown by color contour. Arrows indicate the direction of $v_{\rm pol}$ only, and their length does not reflect the absolute value.}\label{fig:Map-Ma}
\end{figure}

The bottom panels of Figure~\ref{fig:accretion-midplane-surface} show that $\alpha_{\phi z}^{\rm m}$ is much smaller than $\alpha_{R \phi}^{\rm m}$ even around the disk surface, which means that the angular momentum loss by the disk wind is unimportant near the disk surface. The reason for this arises from both our initial condition and the characteristics of the MRI-driven wind. The initial disk field strength is assumed to be weak ($\beta=10^4$) and the disk gas extends well above the disk due to the MRI-driven wind. As a result, the Alfv\'en Mach number, defined by the ratio of the poloidal velocity $v_{\rm pol}$ to the poloidal Alfv\'en velocity $v_{{\it A}, \rm pol}$, is comparable to or larger than unity even around the disk (see Figure~\ref{fig:Map-Ma}). For this reason, a magnetic field cannot behave as a rigid wire nor carry angular momentum efficiently in the vertical direction (see also the bottom right panel in Figure~\ref{fig:accretion-midplane-surface}).

\subsection{Global Velocity and Magnetic Field Structures}
\begin{figure}
\epsscale{1.0}
\plotone{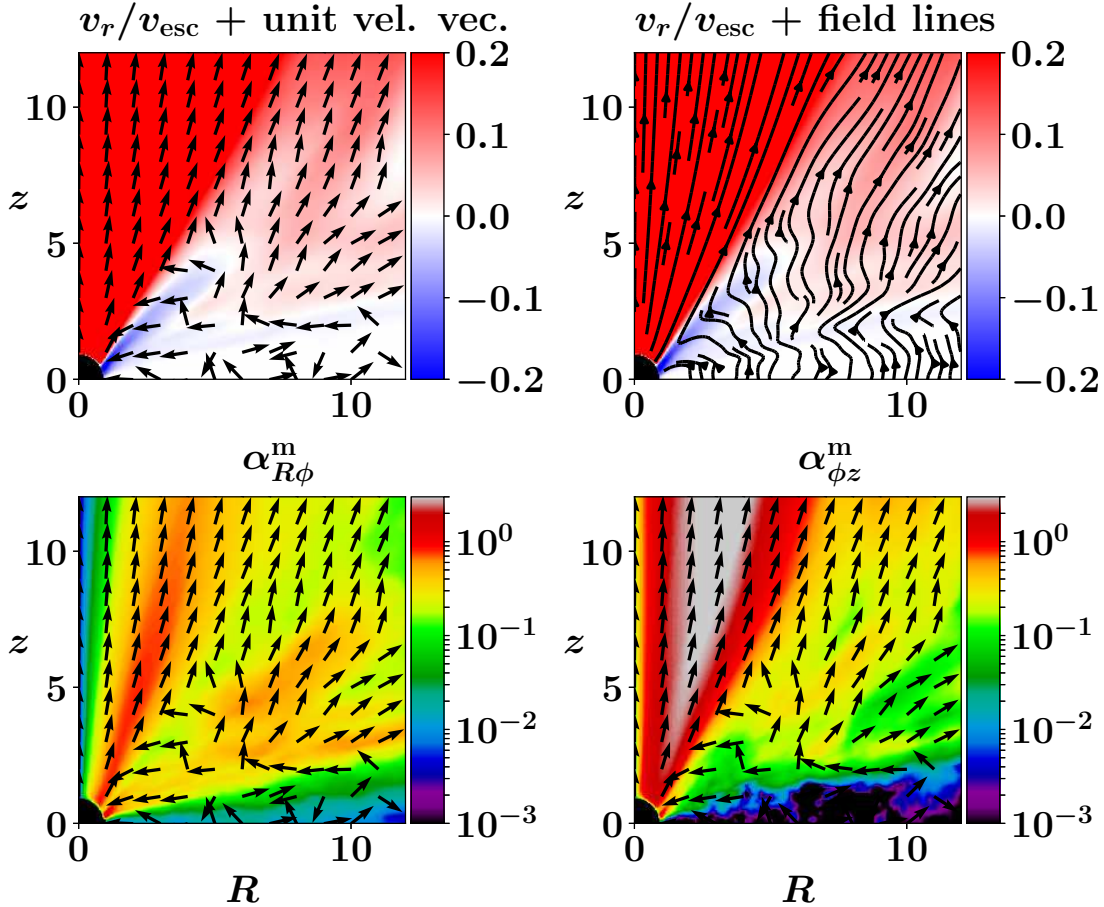}
\caption{Accretion structure near the center. Top panels: the radial velocity normalized by the escape velocity, with unit vectors that show the direction of the poloidal velocity (left) and poloidal field lines (right). Bottom: magnetic $\alpha$ parameters, $\alpha_{R\phi}^{\rm m}$ (left) and $\alpha_{ \phi z}^{\rm m}$ (right). Unit vectors of the poloidal velocity are also shown. All the quantities are temporally and azimuthally averaged. The time average is performed during $t=250t_{\rm K0}$ to $300t_{\rm K0}$.}\label{fig:accretion-structure}
\end{figure}

The averaged velocity field and magnetic field structures are displayed in the top panels of Figure~\ref{fig:accretion-structure}. The top left panel shows that the gas stream from the disk surface roughly splits into two branches: the upflow from the outer disk becomes a weak disk wind, and the wind from the inner disk accretes onto the central star. The wind of the second branch fails to blow out, accreting onto the star with a large velocity. That is, the failed disk wind becomes the funnel-wall accretion. The bottom panels of Figure~\ref{fig:accretion-structure} show that the $\alpha$ parameters due to the Maxwell stress (including the $\phi z$ component, $\alpha_{ \phi z}^{\rm m}$) are close to unity in the upper atmosphere, which means that very efficient angular momentum exchange due to the Maxwell stress causes the funnel-wall accretion. We will study the mechanism of the funnel-wall accretion in more detail in Section~\ref{subsec:funnel}.

The top right panel of Figure~\ref{fig:accretion-structure} also shows that the magnetic fields near the center are dragged toward the star. As shown in Figure~\ref{fig:accretion-rate-structure}, the accretion speed is larger in the upper disk atmosphere than in the disk. The accreting materials above the disk drag magnetic fields toward the star, forming this magnetic structure. The transport of poloidal magnetic fields above the disk has also been discussed previously. The efficient transport in the upper atmosphere was first pointed out by \citet{matsumoto1996}, and termed as the ``coronal mechanism" by \citet{beckwith2009} \citep[see also ][]{zhu2017}. 

\begin{figure}
\epsscale{.50}
\plotone{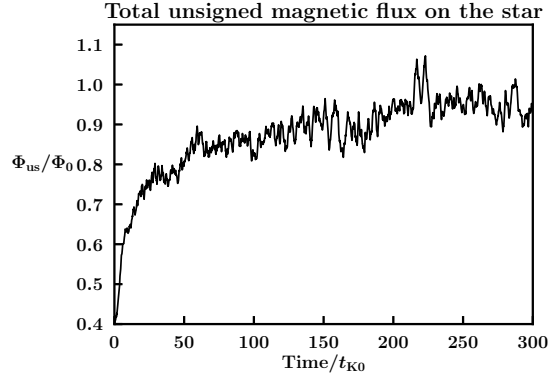}
\caption{Temporal evolution of the total unsigned magnetic flux $\Phi_{\rm us}$ measured on the stellar surface. Also see Figure~\ref{fig:accretion-structure}, where how a poloidal field is carried to the star by accreting material is shown. The unit of time is $t_{\rm K0}$. The unit of the magnetic flux is $\Phi_0 = B_0 L_0^2$.}\label{fig:stellar-magnetic-flux}
\end{figure}

The funnel-wall accretion and the disk surface accretion carry poloidal magnetic fields toward the star. The accumulated magnetic fields form the magnetic funnels around the poles (see the top right panel of Figure~\ref{fig:accretion-structure}). 
A fraction of the magnetic fields of the magnetic funnels are carried from the disk through the coronal mechanism as in simulations of black hole accretion disks \citep{devilliers2003,beckwith2009}. 
Figure~\ref{fig:stellar-magnetic-flux} shows the total amount of the unsigned magnetic flux through the stellar surface ($r=1R_*$). The magnetic flux sharply increases in the early phase, but the rapid accumulation stops after $t\sim 60t_{\rm K0}$. We can translate the total flux into the field strength averaged at the stellar surface by multiplying by $1/4\pi$. Since the total flux is almost 0.9 in our units, the averaged field strength is 72 and 160~G for CTTSs and HAeBes, respectively.  Note that mass accretion onto the star continues even when the magnetic flux becomes almost constant, which is a result of decoupling between a magnetic field and gas due to turbulent diffusion.

\begin{figure}
\epsscale{.90}
\plotone{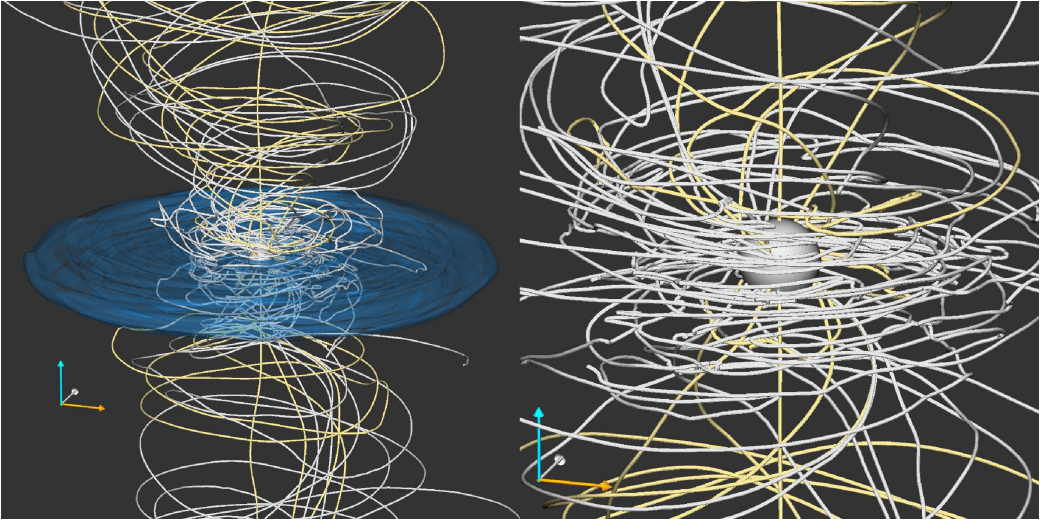}
\caption{3D structure of magnetic field lines at $t=285t_{\rm K0}$. The field lines that pass through the stellar surface are colored yellow. The blue contour in the left panel shows the density isosurface with the density of $0.03\rho_0$. The right panel is an enlarged image of the central region without the isosurface.}\label{fig:fieldline-3D}
\end{figure}

The magnetic fields are significantly bent in the funnel-wall accretion region due to the strong drag, forming a dipole-like shape on this plane (see Figure~\ref{fig:accretion-structure}). However, the 3D structure of the magnetic field is very different from the pure dipole magnetosphere. Figure~\ref{fig:fieldline-3D} displays a snapshot of the 3D structure of the magnetic field. The magnetic field is mainly toroidal around the star because of the differential rotation, which is distinct from the pure dipole magnetosphere.

\subsection{Funnel-wall Accretion onto the Central Star}\label{subsec:funnel}
It has been believed for a long time that a fast accretion to the high-latitude region of the star is a clear sign of the magnetospheric accretion \citep{camenzind1990,koenigl1991}. However, our simulation indicates that a fast, high-latitude accretion is possible even without a stellar magnetosphere. Here we investigate the behavior and the physics of the funnel-wall accretion.

\subsubsection{Structure of Accretion Streams}
\begin{figure}
\epsscale{1.0}
\plotone{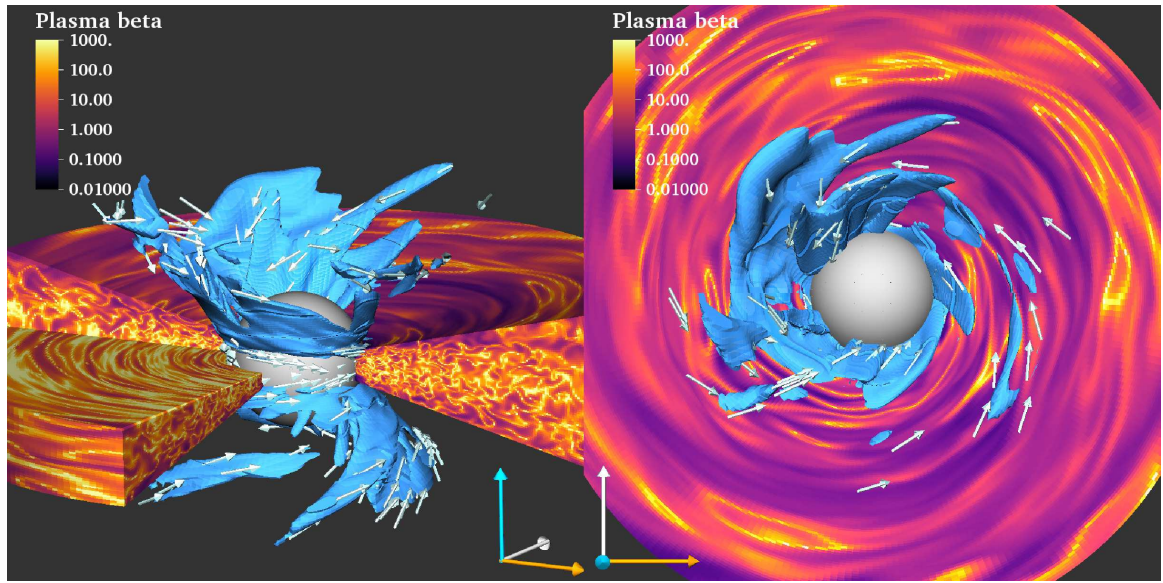}
\caption{Snapshot of funnel-wall accretion flows onto the central star at $t=285t_{\rm K0}$. The central star is shown as the central sphere, and the inner disk is colored with the value of plasma $\beta$. The blue regions indicate the fast accreting material ($v_{\rm r}<-0.2 v_{\rm K0}$). Arrows denote the direction of velocity vectors.}\label{fig:accretion-3D}
\end{figure}

The 3D structure of the accretion flow is shown in Figure~\ref{fig:accretion-3D}. The disk extends to the stellar surface, directly accreting the disk material to the star. 
Therefore, the so-called boundary-layer accretion is taking place \citep[i.e. the inner disk is not truncated by the stellar magnetic field, unlike in the magnetospheric accretion model;][]{popham1993,romanova2012}. 
Simultaneously, multiple fast accretion streams are falling to high-latitude areas of the star. One will notice that the accretion streams start well above the disk, which is different from the previously reported disk surface accretion \citep{suzuki2014}.

\begin{figure}
\epsscale{.60}
\plotone{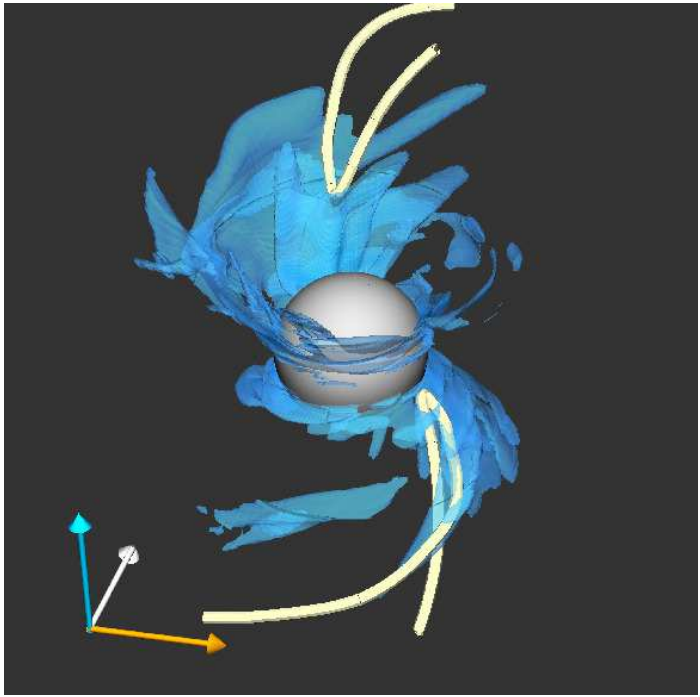}
\caption{3D structure of magnetic fields threading accretion streams at $t=285t_{\rm K0}$. The blue regions indicate the fast accreting material ($v_{ r}<-0.2 v_{\rm K0}$). Two lines denote magnetic field lines dragged by accretion matter.}\label{fig:BentFieldLines}
\end{figure}

Figure~\ref{fig:BentFieldLines} displays the structure of the magnetic field threading fast accretion streams. The magnetic field is strongly bent owing to the dragging of the accreting matter. Unlike the magnetospheric accretion model, accretion flows are not guided by the magnetic field.

\begin{figure}
\epsscale{.80}
\plotone{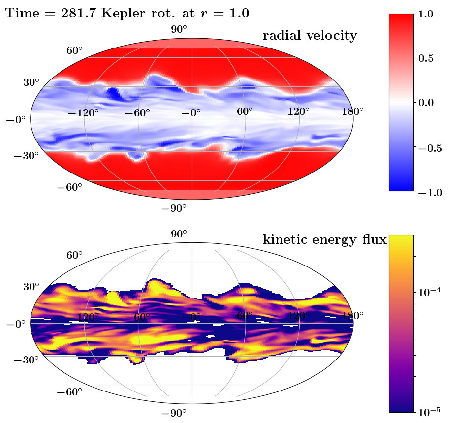}
\caption{Mollweide projection of the radial velocity (top) and accreting kinetic energy flux ($-\rho v_{ r}^3$, bottom) distributions on the stellar surface ($r=1R_*$) at $t=281.7t_{\rm K0}$. Note that the polar outflowing regions indicate the stellar wind regions. An animation of this figure is available.}\label{fig:vrad-moll}
\end{figure}

The fragmented accretion streams form patchy accretion spots on the stellar surface. The top panel of Figure~\ref{fig:vrad-moll} exhibits the Mollweide projection of the radial velocity distribution on the stellar surface ($r=1R_*$). The accretion speed is small near the disk midplane. However, above $\pm \sim 30^{\circ}$ from the midplane, the gas is falling with almost the Keplerian velocity on the stellar surface. Since this velocity is supersonic, accretion shocks are expected. The accretion spot pattern largely varies with time, but patchy accretion spots with a large velocity are always present. The bottom panel displays the kinetic energy flux of the accreting matter ($\rho v_{ r}^3$, only the regions with $v_{ r}<0$ are visualized), which shows that the fast accreting matter injects a large kinetic energy flux. Regions with a large kinetic energy flux could be observed as hot spots.

\begin{figure}
\epsscale{.50}
\plotone{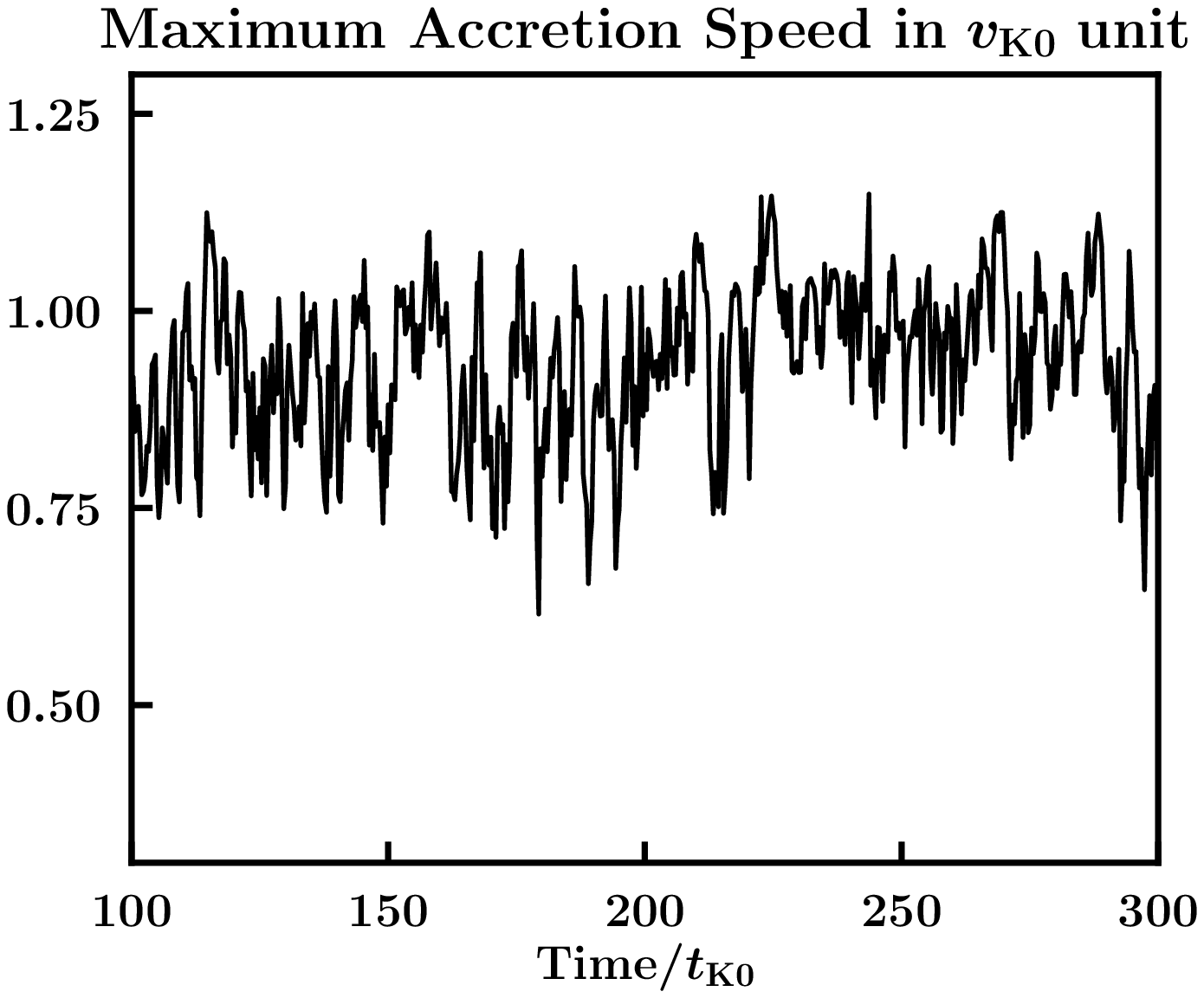}
\caption{Temporal evolution of the maximum accretion velocity measured at the stellar surface. The velocity is normalized by the Keplerian velocity at the stellar surface. The unit of time is $t_{\rm K0}$.}\label{fig:max-accvel}
\end{figure}

Figure~\ref{fig:max-accvel} shows the temporal evolution of the maximum accretion velocity measured at the stellar surface. The plot indicates that the maximum speed is highly time-variable but is typically close to the Keplerian velocity at the stellar radius. The variability originates from the MRI turbulence. The accretion speed is moderately slower than the escape velocity (the accretion velocity is $\sim v_{\rm K}$, 70~\% of the escape velocity). 

\begin{figure}
\epsscale{1.0}
\plotone{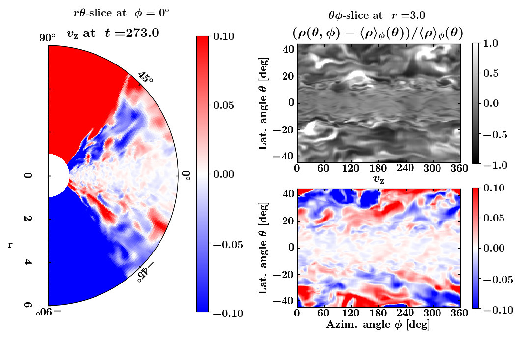}
\caption{Intermittency of the MRI-driven wind. Left: a snapshot of $v_{z}$ map on the $r \theta$ slice plane at $t=273t_{\rm K0}$ (note that the data are neither azimuthally nor temporally averaged). Right: density fluctuation map (top)  and $v_{z}$ (bottom) on the $\theta \phi$ plane at $r=3R_*$ and $t=273t_{\rm K0}$. Note that upflows from and downflows toward the disk surface are mixed up, and the density fluctuation is highly inhomogeneous in both the azimuthal and latitudinal directions. An animation of this figure is available. The sequence starts at time 200$t_{\rm K0}$ and ends at time 300$t_{\rm K0}$. The animation duration is 8~s.}\label{fig:slice_drho_vz}
\end{figure}

The spatiotemporal intermittency of the funnel-wall accretion arises from the intermittent nature of the MRI-driven wind. The left panel of Figure~\ref{fig:slice_drho_vz} displays the vertical gas motion on the $r\theta$ plane. Since the data are averaged neither azimuthally nor temporally, we can see the fine velocity structure in this figure. We can find that upflows from and downflows toward the disk surface are mixed up. The mixture is a characteristic feature of the MRI-driven wind. The bottom right panel shows the slice of $v_{ z}$ on the $\theta \phi$ plane at $r=3R_*$ at the same time, which indicates that the mixture of upflows and downflows is also evident in the azimuthal direction. The density fluctuation in the top right panel is also highly inhomogeneous in both the azimuthal and latitudinal directions.

\begin{figure}
\epsscale{.50}
\plotone{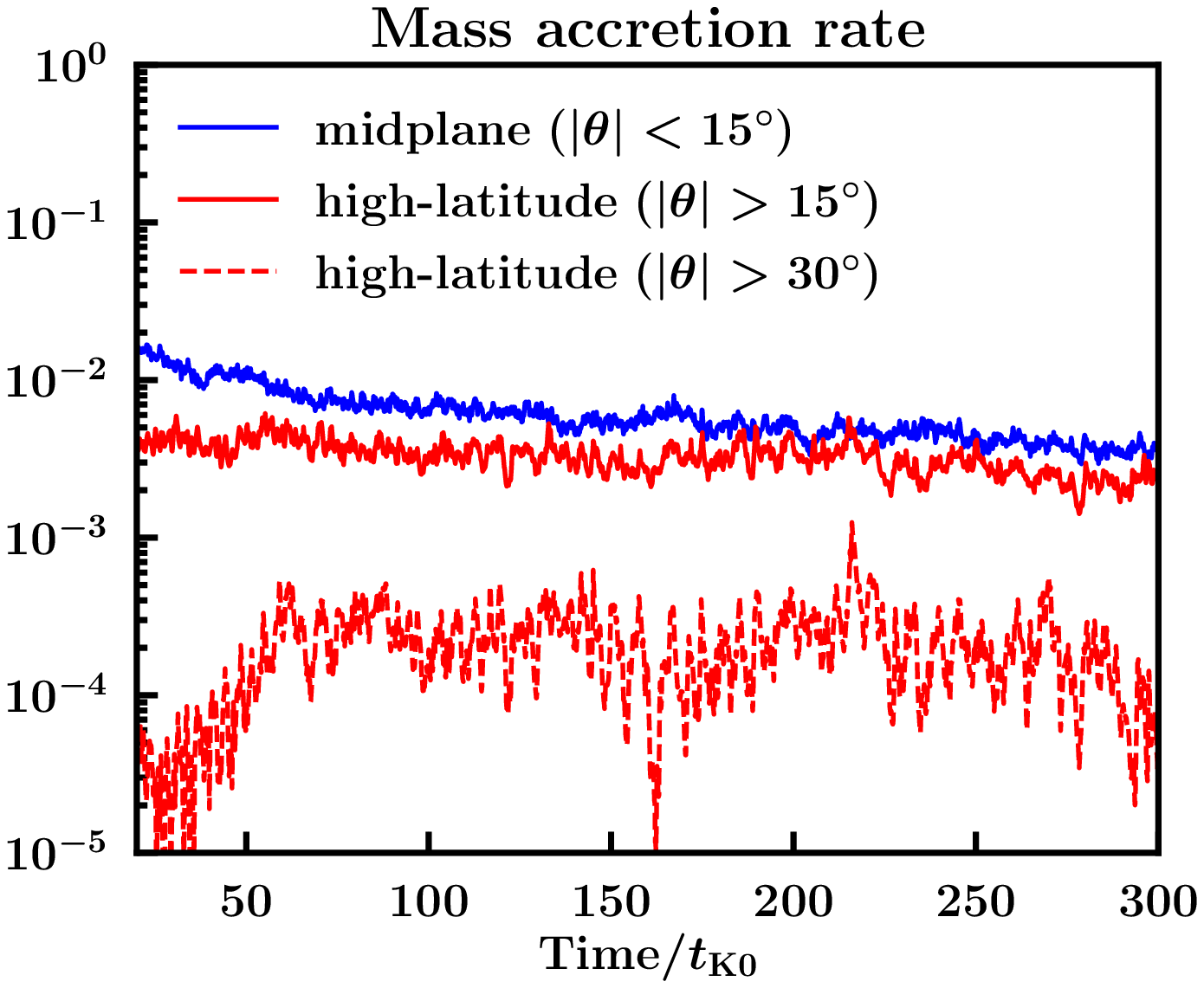}
\caption{Mass accretion rates measured in the latitudinal ranges of $|\theta|< 15^{\circ}$ (blue solid line), $|\theta|> 15^{\circ}$ (red solid), and $|\theta|> 30^{\circ}$ (red dashed). The angle of $15^\circ$ almost corresponds to the height of $z=2H_{\rm p}$. The unit of time is $t_{\rm K0}$.}\label{fig:mdot}
\end{figure}

As shown in Figures~\ref{fig:overview} and \ref{fig:accretion-3D}, the inner disk reaches the stellar surface in our model. In Figure~\ref{fig:mdot}, we compare the accretion rate through the region near the disk midplane (within $\pm 15^{\circ}$, which almost corresponds to the height of $|z|=2H_{\rm p}$.) and the accretion rate onto the high-latitude areas. Since it is difficult to clearly define the latitudinal domain into the disk and high-latitude regions separately, we plot the two curves for the accretion rate onto the high-latitude areas for reference. One is measured in the regions with $|\theta|>15^\circ$ (red solid line, representing the accretion rate measured above the typical disk height, $|z|=2H_{\rm p}$), and the other is in the regions with $|\theta|>30^\circ$ (red dashed). The former is similar to or a factor of 2 smaller than the accretion rate of the midplane, while the latter is several tens of times smaller than that. Therefore the mass accretion rate of the funnel-wall accretion well above the disk is much smaller than that of the disk accretion. If we take the normalization units for CTTSs, the total accretion rate is $\sim10^{-9}~M_{\odot}~{\rm yr}^{-1}$, which is comparable to the typical observed accretion rate.

\subsubsection{Role of Disk Dynamo}
We have seen that the angular momentum exchange due to the Maxwell stress in the upper atmosphere is the key for the formation of the funnel-wall accretion (e.g. Figure~\ref{fig:accretion-structure}). The magnetic fields in the MRI-turbulent disk are time-variable and have a complex structure, which requires us to carefully see behaviors of the magnetic fields such as amplification and transport. Here we investigate behaviors of the magnetic fields near the star.

\begin{figure}
\epsscale{1.0}
\plotone{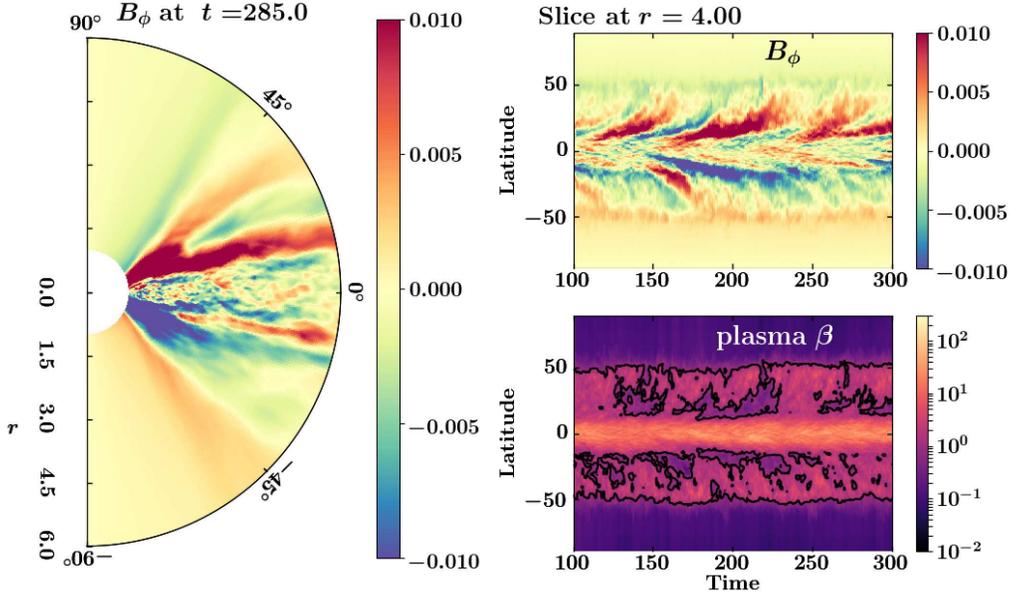}
\caption{Dynamo activity of the disk. Left: snapshot of the azimuthally averaged toroidal field $B_\phi$ map, Right: time-latitude diagrams of $B_\phi$ (top) and plasma $\beta$ (bottom). The time-latitude diagrams are created using the values at $r=4R_*$. The contour in the plasma $\beta$ map indicates the plasma $\beta = 1$. The unit of time is $t_{\rm K0}$. An animation of this figure is available. The sequence starts at time zero and ends at time 317$t_{\rm K0}$. The animation duration is 25~s.}\label{fig:dynamo-Bphi-beta}
\end{figure}

MRI disks show dynamo activities in which magnetic fields are amplified by shearing motions \citep[e.g.][]{stone1996,hawley1996}. Figure~\ref{fig:dynamo-Bphi-beta} displays the dynamo activity in our simulation. The left panel shows a snapshot of the azimuthally averaged toroidal field $B_\phi$ map at $t=285t_{\rm K0}$. One can see that a toroidal field is mostly amplified around the disk surface. To see the temporal evolution, we made the time-latitude diagrams of $B_\phi$ and the plasma $\beta$ (right panels), measured at a radius of $r=4R_*$. The time-latitude diagram of $B_\phi$ shows a so-called butterfly diagram in which the toroidal field reverses its sign quasi-periodically, as in previous studies \citep[e.g.][]{gressel2010a,machida2013}.

Unlike in the local shearing box simulations, the positive and negative $B_{\phi}$ are dominant in the upper and lower hemispheres, respectively, which is an indication of the global effects \citep[see also][]{suzuki2014}. We have seen that magnetic fields above the disk are dragged toward the star. This results in the creation of a global, negative $B_{ R}$ in the upper hemisphere. The radial differential motion converts the negative $B_{R}$ into the positive $B_{\phi}$. Since the global $B_{R}$ values always exist, the symmetry in the radial direction breaks, and therefore the positive $B_{\phi}$ becomes dominant in the upper hemisphere. The same process operates in the lower hemisphere. The occurrence of the sign reversal of the toroidal field depends on the competition between the amplification of the global fields and the local dynamo; we see the reversal around $r=4R_*$, but the reversal disappears within $r\lesssim 2R_*$.

The amplified magnetic fields escape from the disk due to magnetic buoyancy \citep[Parker instability;][]{parker1955,parker1966,miller2000,shi2010}. Strongly magnetized regions ($\beta\sim 1$) are formed near the disk surface, but the growth timescale of the Parker instability becomes comparable to the local rotational period when $\beta \sim 1$. As a result, the strong magnetic fields erupt from the disk without further amplification. This is the reason why the plasma $\beta$ in the upper atmosphere cannot be much lower than unity even though a magnetic field is continuously supplied from the disk. The eruption due to buoyancy is a key process that transports a magnetic field to the upper atmosphere.

\begin{figure}
\epsscale{1.2}
\plotone{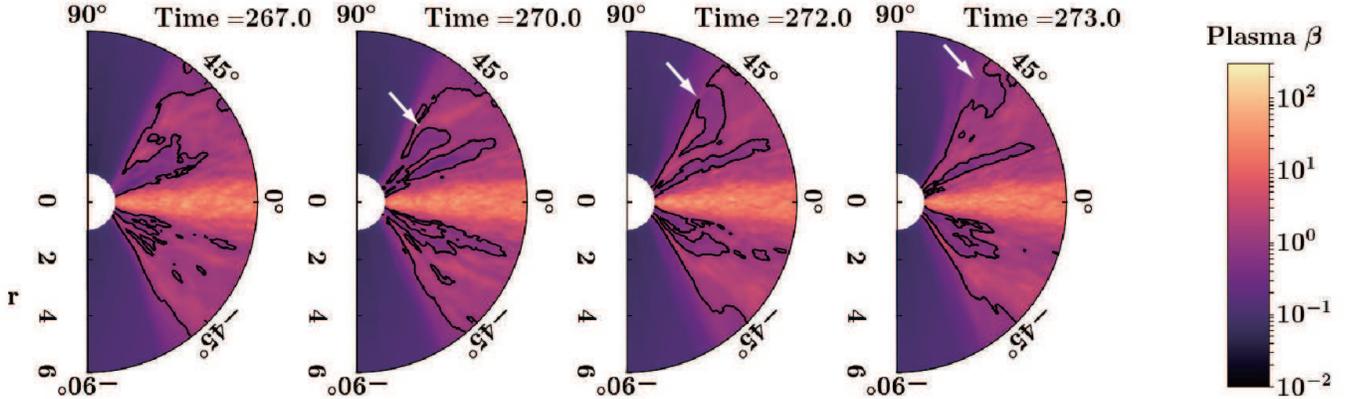}
\caption{Escape of a buoyant magnetic flux bundle. A low-$\beta$ region (indicated by the white arrows) is moving outward as time progresses. The black contours indicate the plasma~$\beta=1$. The unit of time is $t_{\rm K0}$.}\label{fig:beta-voids}
\end{figure}

Figure~\ref{fig:beta-voids} presents an example of erupting magnetic flux bundles. The time proceeds from left to right. At $t=267t_{\rm K0}$, a low-$\beta$ region is formed near the disk surface. As time progresses, the low-$\beta$ void erupts upward. Note that the movement of magnetic fields and the movement of plasmas are opposite here: the erupting magnetic fields are moving outward, while plasmas are accreting onto the star. We find that the ejection of low-$\beta$ voids occurs recurrently. In addition, the low-$\beta$ flux bundles break up and mix with the surrounding higher-$\beta$ plasma during the ascending motion.

\begin{figure}
\epsscale{.80}
\plotone{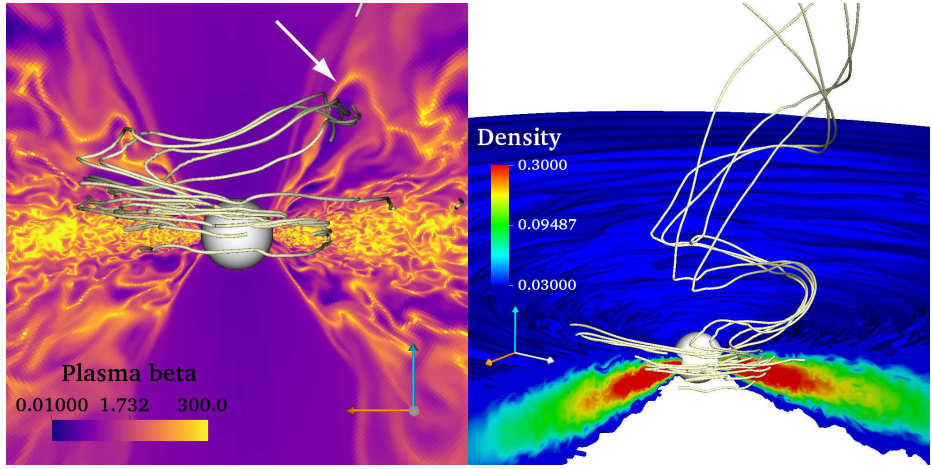}
\caption{Eruption of a magnetic flux bundle from the disk at $t=272t_{\rm K0}$. Left: the central sphere denotes the star. Field lines that penetrate a rising, low-$\beta$ region are shown. The 2D slice of the plasma $\beta$ is also displayed. The low-$\beta$ void indicated by the white arrow corresponds to the buoyant magnetic flux bundle in Figure~\ref{fig:beta-voids}. Right: a bird's-eye view of the erupting field lines plotted in the left panel. The disk is colored by the density value.}\label{fig:flux-bundle-ejection}
\end{figure}

The 3D structure of the erupting magnetic flux bundle in Figure~\ref{fig:beta-voids} is shown in Figure~\ref{fig:flux-bundle-ejection}. We visualize field lines that penetrate the erupting, low-$\beta$ void above the disk (upper right from the star, in the left panel). As expected, those field lines shape an undulating flux bundle (left panel). The one end of the lifted part is anchored in the disk, but the other end is erupting upward (right panel). This figure demonstrates how buoyant magnetic flux bundles erupt from the disk.

\begin{figure}
\epsscale{.50}
\plotone{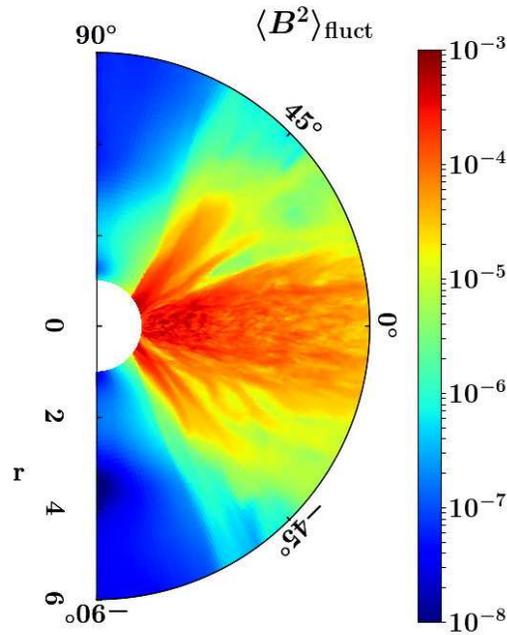}
\caption{Snapshot of the fluctuating component of magnetic energy at $t=270t_{\rm K0}$. The fluctuating component $\langle B^2\rangle_{\rm fluct}$ is defined as $\langle B^2\rangle_{\rm fluct}=\langle B^2 \rangle_{\phi} - \langle B \rangle_{ \phi}^2.$}\label{fig:B2-F-around-star}
\end{figure}

The magnetic fields erupting from the inner disk move along the magnetic funnels around the poles (Figure~\ref{fig:beta-voids}), which suggests that the magnetic fields are strong along the funnels. Figure~\ref{fig:B2-F-around-star} supports this speculation, where the fluctuating component of magnetic energy is shown. The fluctuating component is closely related to the erupting magnetic flux bundles, because the erupting bundles have a highly nonuniform structure in the azimuthal direction. This magnetic field transport occurs recurrently and leads to the enhancement of the angular momentum loss around the magnetic funnels.

\begin{figure}
\epsscale{.60}
\plotone{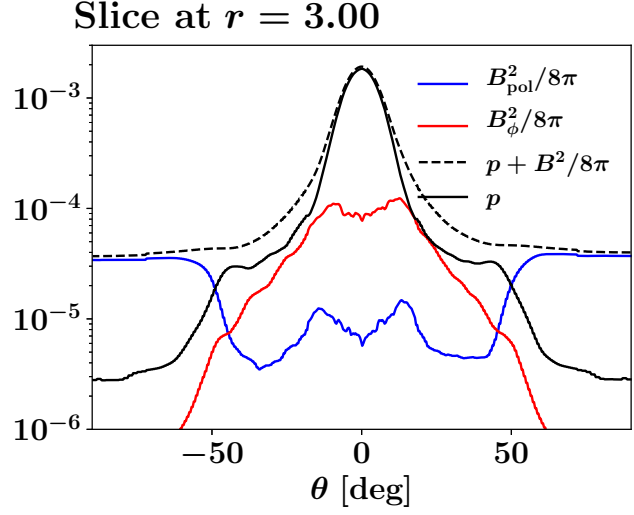}
\caption{Latitudinal pressure balance. Here, $\theta=0^\circ$ corresponds to the midplane, while $\theta=\pm90^\circ$ corresponds to the north/south poles. Note that around $\theta=\pm 50^\circ$ the gas pressure on the disk side balances with the magnetic pressure of the poloidal field around the poles. The data are temporally averaged during the time $t=250t_{\rm K0}$ to $300t_{\rm K0}$.}\label{fig:pressure-balance-theta}
\end{figure}

The opening angle of the magnetic funnels determines the maximum latitude of the funnel-wall accretion (see Figure~\ref{fig:accretion-structure}). To see how the angle is determined, we examine the pressure balance in the latitudinal direction. Figure~\ref{fig:pressure-balance-theta} shows the distributions of different kinds of pressure against the latitude at $r=3R_*$. From the midplane ($\theta=0^\circ$) to a certain height ($\theta\sim \pm 50^\circ$), the gas pressure is the most dominant term. However, the magnetic pressure of a poloidal field is the largest around the poles and balances with the gas pressure on the disk side at $\theta\sim \pm 50^{\circ}$. The disk materials and therefore erupting magnetic fields cannot easily go beyond those angles because of the magnetic pressure walls (funnel walls).

\begin{figure}
\epsscale{.50}
\plotone{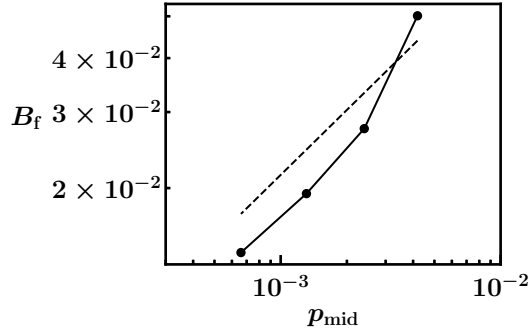}
\caption{Numerical check of Equation~\ref{eq:funnel-balance}, which indicates that the magnetic field strength of the magnetic funnel is determined by the disk gas pressure. The solid circles from right to left are the data points at $r=2R_*$, $3R_*$, $4R_*$, and $5R_*$. The dashed line denotes $B_{\rm f}=\sqrt{8\pi e^{-4}p_{\rm mid}}$.}\label{fig:B-funnel}
\end{figure}

We will show that the field strengths in the magnetic funnels are determined by the disk gas pressure. The gas pressure in our disk can be expressed as $p(R,z) = p_{\rm mid}(R)\exp{\left [-(z/H_{\rm p})^2 \right]}$ in the disk, where $p_{\rm mid}$ is the pressure at the midplane. Since the gas pressure distribution takes a flat profile above the disk surface $z=2H_{\rm p}$, we approximate that the gas pressure above the disk $p_{\rm up}$ is $p_{\rm up} = p(R,2H_{\rm p})= p_{\rm mid}(R)e^{-4}$. Therefore, if we consider the pressure balance at the position $(R_{\rm f},z_{\rm f})$, we obtain the following relation between the disk pressure and the field strength $B_{\rm f}$ (the subscript ``f" denotes the magnetic funnel):
\begin{align}
\frac{B_{\rm f}(R_{\rm f},z_{\rm f})^2}{8\pi}&\approx p_{\rm up}(R_{\rm f},z_{\rm f})\\
B_{\rm f}(R_{\rm f},z_{\rm f}) &\approx \sqrt{8\pi e^{-4} p_{\rm mid}(R_{\rm f})}.\label{eq:funnel-balance}
\end{align}
Figure~\ref{fig:B-funnel} examines this relation in our simulation using the data at $r=2R_*$, $3R_*$, $4R_*$, and $5R_*$. We first measure the cylindrical radius and $B_{\rm f}$ at the balance points at these radii, and we then obtain $p_{\rm mid}$ at the corresponding cylindrical radii. This figure indicates that this relation holds fairly well.

\subsubsection{Angular Momentum Exchange Process}\label{sec:ang-mom-exchange}
We have seen that the funnel-wall accretion originates from the failed disk wind that loses angular momentum and is trapped by the stellar gravitational potential. We investigate the angular momentum exchange process in the disk wind.

\begin{figure}
\epsscale{1.0}
\plotone{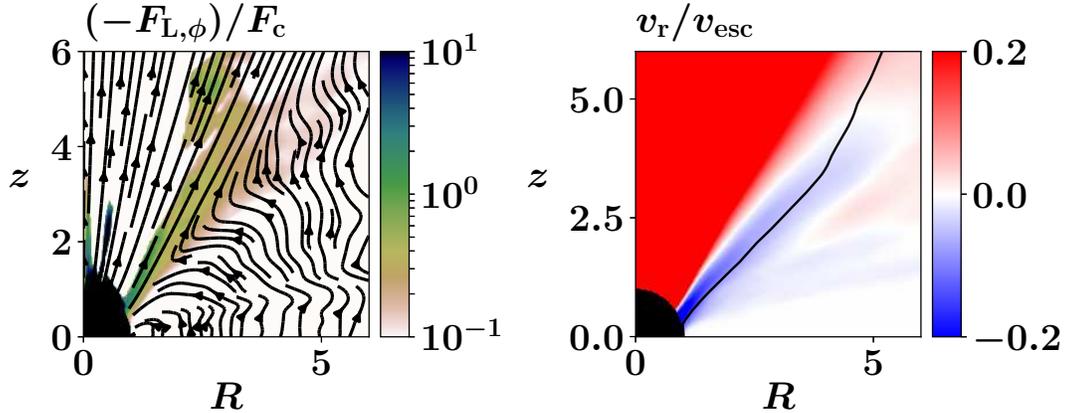}
\caption{Left: the relative magnitude between the azimuthal component of the Lorentz force and the centrifugal force. Lines show magnetic field lines. Right: radial velocity normalized by the local escape velocity. The solid line indicates the position where the specific angular momentum is $R_{*} v_{\rm K0}=1$ and denotes the centrifugal barrier. In both panels, the quantities are azimuthally and temporally averaged.}\label{fig:FL-Fc}
\end{figure}

A rapid angular momentum loss is necessary for the generation of the nearly free-fall accretion. In order for the angular momentum loss to occur on an orbital timescale, the Lorentz force that decelerates the rotating motion of the matter should be comparable to the centrifugal force \citep{matsumoto1996}. Figure~\ref{fig:FL-Fc} compares the spatial distribution of the ratio of the two forces ($(-F_{\rm L,\phi})/F_{\rm c}$, the minus sign is added because we focus on the decelerating force) and the radial velocity structure. The region with the negative radial velocity almost coincides with the region with $(-F_{\rm L,\phi})/F_{\rm c}\gtrsim 0.3$, and is along the magnetic funnel. As seen in Figures~\ref{fig:beta-voids} and \ref{fig:B2-F-around-star}, the magnetic field strength along the funnel is larger than that in the surrounding regions because of the magnetic field transported from below. This explains why the rapid angular momentum loss occurs around the magnetic funnel. We note that the material around the starting point of the funnel-wall accretion ($(R,z)\approx (5R_*,5R_*)$) is magnetically disconnected from the star, unlike in the magnetospheric accretion model. Therefore, the effect of the inner boundary condition should be insignificant for the formation of the funnel-wall accretion.

The right panel of Figure~\ref{fig:FL-Fc} shows the location where the specific angular momentum is the same as that of a gas rotating at the stellar surface with the Keplerian velocity (see also the fourth panel of Figure~\ref{fig:accretion-structure}). The isosurface is radially inclined away from the poles because the radial angular momentum transport is more efficient in the upper atmosphere than in the lower atmosphere. Once a material comes inside the cone of the isosurface, the material becomes rotationally unsupported, which leads to the formation of the fast accretion.

Figure~\ref{fig:fieldline-3D} shows that the angular momentum exchange in the funnel-wall accretion is mediated by magnetic fields that connect a rapidly rotating inner material with a slowly rotating outer one. This mechanism is general for rotating systems and is basically the same as the physics that operates at the onset of MRI and in the coronal mechanism and the magnetic braking.

We derive a useful indicator for the occurrence condition of the funnel-wall accretion. Since the plasma $\beta$ is close to unity in the region of interest, we expect that $v_{\it A}\approx c_{\rm s}$. For the generation of the nearly free-fall accretion, the timescale of the angular momentum exchange should be at most the orbital timescale $t_{\rm orb}$. This defines the length scale for the angular momentum transport: $L=v_{\it A} t_{\rm orb}\approx c_{\rm s}/\Omega^\prime$, where $\Omega^\prime = v_{\phi}/R$ is the local angular velocity. The expression of $L$ is similar to the standard definition of the disk pressure scale height, but defined well above the disk. We can estimate the magnetic tension force exerted from bent magnetic fields as $B^2/(4\pi \rho R_{\rm curv})$, using a typical length scale for the curvature radius $R_{\rm curv}$. If we replace $R_{\rm curv}$ with the angular momentum exchange length scale $L$, we arrive at the following expression of the relative magnitude of the Lorentz force $F_{\rm L,\phi}$ and the centrifugal force $F_{\rm c}$:
\begin{align}
\frac{|F_{\rm L,\phi}|}{F_{\rm c}} \approx \frac{B^2}{4\pi \rho L}\frac{R}{v_\phi^2} \approx \frac{v_{\it A}}{c_{\rm s}}\frac{v_{\it A}}{v_{\phi}}\approx \left( \beta \beta_{\rm rot}\right)^{-1/2},
\end{align}
where $\beta_{\rm rot}\equiv (v_{\phi}/v_{\it A})^2$ represents the ratio of the kinetic energy density of the rotational motion to the magnetic energy density. When $\left( \beta \beta_{\rm rot}\right)^{-1/2}$ is close to unity, we expect a rapid angular momentum loss on a timescale of the local rotation.

\begin{figure}
\epsscale{1.0}
\plotone{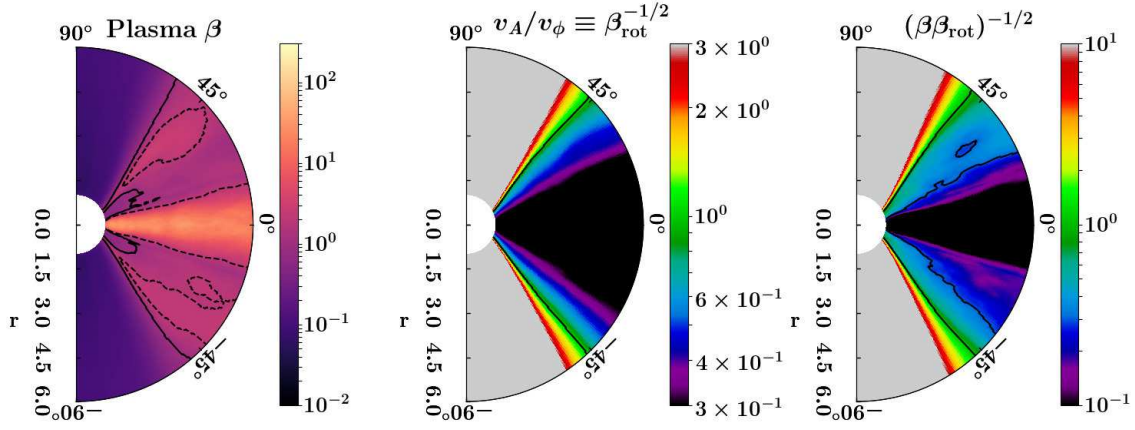}
\caption{Examination of the indicator of the occurrence condition of the funnel-wall accretion. Left: plasma $\beta$. The solid and dashed lines indicate isosurfaces with $\beta=1$ and $2$, respectively. Middle: $v_{\rm A}/v_{\phi}=\beta_{\rm rot}^{-1/2}$. The line denotes the location where $\beta_{\rm rot}=1$. Right: the indicator $\left(\beta \beta_{\rm rot}\right)^{-1/2}$. Solid lines indicate the locations where $\left(\beta \beta_{\rm rot}\right)^{-1/2}=0.3$ and $1$.}\label{fig:beta-betarot}
\end{figure}

Figure~\ref{fig:beta-betarot} examines the above discussion about the occurrence condition of the funnel-wall accretion. The left panel exhibits that the plasma $\beta$ is approximately unity in the upper disk atmosphere. The middle panel indicates $v_{\it A}/v_{\phi}=\beta_{\rm rot}^{-1/2}$, which shows that the magnetic energy density increases and becomes comparable to the kinetic energy of the rotational motion with increasing latitude. The right panel displays the indicator of the occurrence condition $\left( \beta \beta_{\rm rot}\right)^{-1/2}$. This figure demonstrates that the indicator is indeed close to unity around the regions of the funnel-wall accretion.

\begin{figure}
\epsscale{1.0}
\plotone{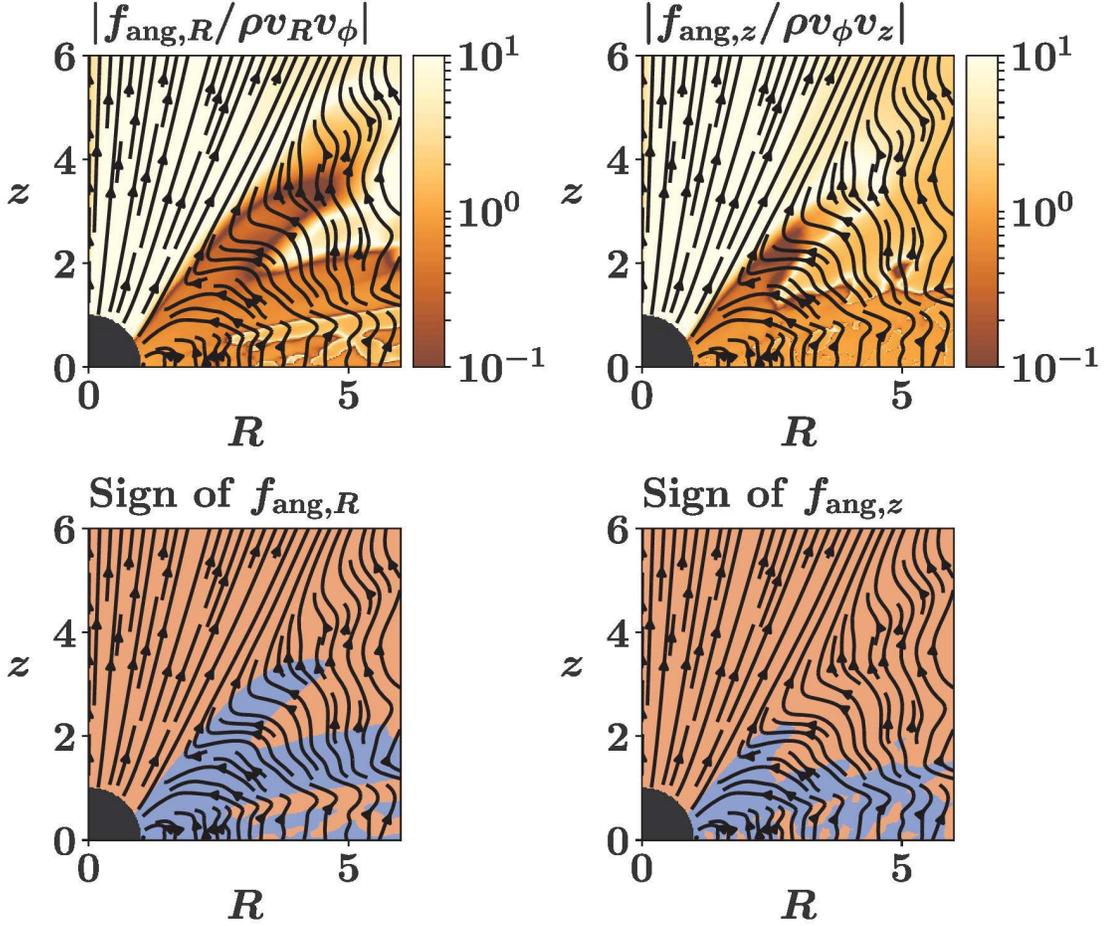}
\caption{Angular momentum flux around the star. The top panels show the magnitudes of the angular momentum flux in the $ R$ (left) and $z$ (right) directions. The values are normalized by the angular momentum flux carried by the fluid. The bottom panels indicate the sign of the angular momentum shown in the top panels, where the red and blue colors denote the positive and negative signs, respectively. The solid lines show averaged poloidal magnetic fields projected onto this plane. The data is temporally averaged during the time $t=250t_{\rm K0}$ to $300t_{\rm K0}$.}\label{fig:angmom-flux}
\end{figure}

We further investigate the angular momentum transport in more detail, which is important for understanding the reason why the funnel-wall accretion is so fast. Figure~\ref{fig:angmom-flux} evaluates the angular momentum flux around the star. The total angular momentum flux in the $R$ and $z$ directions are expressed as 
\begin{align}
f_{\rm ang,R} & = \langle \rho v_{R}v_{\phi} \rangle_{\phi,t} -  \frac{\left \langle B_{ R}B_{\phi}\right \rangle_{\phi,t}}{4\pi}, \\
f_{\rm ang,z} & = \langle \rho v_{z}v_{\phi} \rangle_{\phi,t} -  \frac{\left \langle B_{ z}B_{ \phi}\right\rangle_{\phi,t}}{4\pi},
\end{align}
respectively. The top panels of this figure show the magnitude of these total fluxes normalized by the flux carried by the fluid. The bottom panels display the sign of these total fluxes. The starting point of the funnel-wall accretion is located around $(R,z)=(5R_*,5R_*)$. The rapid outward transport occurs around this region ($|f_{\rm ang,R}/\rho v_{R}v_{\phi}|\sim |f_{\rm ang,z}/\rho v_{ z}v_{\phi}| \sim 3-5$), which indicates that the angular momentum is rapidly removed by the magnetic field. As a result, the materials there start to fall onto the star, which means that the flux carried by the materials $\rho v_{i}v_\phi$ becomes negative ($i=R,z$). However, the flux carried away by magnetic fields $-B_{i}B_\phi/4\pi$ is positive because of its global configuration around the star. Figure~\ref{fig:angmom-flux} shows that the normalized total flux is well below unity in the funnel-wall region. Therefore, two fluxes are almost balanced in the funnel-wall accretion region; the angular momentum of the accreting materials is continuously removed by the magnetic fields in an efficient way.

The analysis above shows that the normalized total angular momentum flux is well below unity. From this, we obtain the following relation in the funnel-wall accretion:
\begin{align}
\rho v_{\rm acc}v_{\phi} \approx \frac{B_{\rm pol}B_{ \phi}}{4\pi},
\end{align}
where $v_{\rm acc}$ is the accretion speed and $B_{\rm pol}$ is the poloidal magnetic field strength. We rewrite this relation as follows:
\begin{align}
v_{\rm acc} &\approx \frac{B_{\rm pol}B_{\phi}}{4\pi p}\frac{p}{\rho v_\phi}\\
& \approx \alpha_{{\rm pol},\phi }^{\rm m} \frac{c_{\rm s}^2}{v_{\rm K}} = \alpha_{{\rm pol},\phi}^{\rm m} \left( \frac{c_{\rm s}}{v_{\rm K}} \right)^2 v_{\rm K},
\end{align}
where we replace $B_{\rm pol}B_{\phi}/4\pi p$ with $\alpha_{{\rm pol},\phi}^{\rm m}=\sqrt{(\alpha_{R,\phi}^{\rm m})^2+(\alpha_{\phi,z}^{\rm m})^2}$, and approximate $v_{\phi}$ with $v_{\rm K}$. 
The ratio $c_{\rm s}/v_{\rm K}$ is much smaller than unity in the disk because of its low temperature. However, the temperature above the disk is much higher than that in the disk due to the heating by turbulent dissipation, and therefore the ratio $c_{\rm s}/v_{\rm K}$ becomes closer to unity above the disk. Taking the azimuthally and temporally averaged values (the value of $\alpha_{{\rm pol},\phi}^{\rm m}$ is taken from Figure~\ref{fig:accretion-structure}), we obtain
\begin{align}
v_{\rm acc} \approx 0.1 \left( \frac{\alpha_{{\rm pol},\phi}^{\rm m}}{0.5}\right) \left( \frac{c_{\rm s}/v_{\rm K}}{0.5}\right)^2 v_{\rm K},
\end{align}
which is almost consistent with the averaged speed of the funnel-wall accretion. The maximum accretion speed becomes comparable to the Keplerian speed, since $\alpha$ locally takes a larger value (and often exceeds unity). Thus, the rapid angular momentum transport in the upper atmosphere explains the nearly free-fall accretion.
The angular momentum transport in the upper atmosphere is involved with the large-scale magnetic fields and is not a viscous process. The fast accretion becomes possible because the funnel-wall accretion occurs in the hotter region with the larger Maxwell stress than in the disk.

As shown in Figure~\ref{fig:angmom-flux}, the materials in the funnel-wall accretion flows give their angular momentum to outer plasmas. As a result, a wind is driven near the funnel-wall accretion flows (e.g. Figures~\ref{fig:accretion-structure} and \ref{fig:slice_drho_vz}). The maximum wind speed is time-variable as the funnel-wall accretion is and reaches a few 10~\% of the Keplerian velocity at the stellar surface within $r\lesssim 20 R_*$.

\subsection{Summary of the Driving Mechanism of the Funnel-wall Accretion}
\begin{figure}
\epsscale{.80}
\plotone{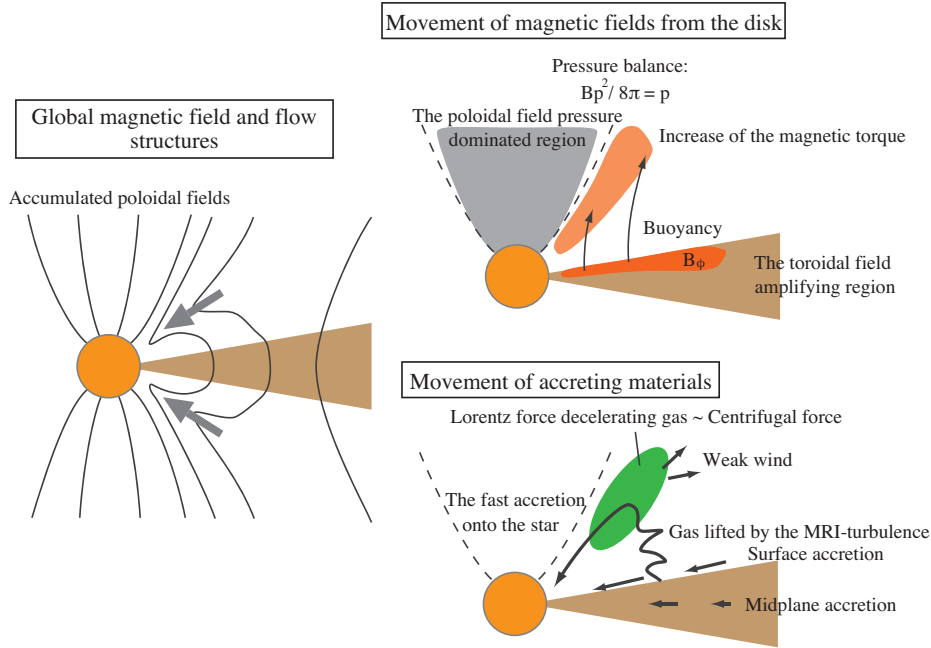}
\caption{Schematic diagrams of accretion and magnetic field structures.}\label{fig:schematic-diagram-summary}
\end{figure}

Figure~\ref{fig:schematic-diagram-summary} summarizes the processes that drive the funnel-wall accretion.
We have seen that the disk material is lifted to the upper atmosphere with the MRI-driven wind (Figures~\ref{fig:overview} and \ref{fig:vertical-distribution}). The materials in the disk wind emanating from the inner disk tend to move inward because of efficient angular momentum loss in the upper atmosphere (Figure~\ref{fig:accretion-structure}). When the gas arrives around the magnetic funnels where amplified magnetic fields are supplied from the inner disk through the Parker instability (Figures~\ref{fig:dynamo-Bphi-beta} to \ref{fig:B2-F-around-star}), the gas experiences a strong torque from the Lorentz force comparable to the centrifugal force (Figures~\ref{fig:FL-Fc} and \ref{fig:beta-betarot}). The angular momentum is rapidly extracted at a place distant from the star. For this reason, when materials lose their angular momenta far enough away from the star, their infall velocity can be comparable to the escape velocity. In this way, various kinds of physics in and above the accretion disk are involved in the formation of the funnel-wall accretion. We also note that a weak wind is driven around the funnel-wall accretion region because of the back reaction of the angular momentum transport (bottom right panel of Figure~\ref{fig:schematic-diagram-summary}; also see Figures~\ref{fig:accretion-structure} and \ref{fig:angmom-flux}). The driving mechanism is essentially the same as the ``micro" Blandford-Payne mechanism proposed by \citet{yuan2012}. We described the process in detail in Section~\ref{sec:ang-mom-exchange}.


\section{Discussion}\label{sec:discussion}
It has been believed for a long time that the magnetospheric accretion is the most promising mechanism that occurs in CTTS because the model predicts the accretion shocks that produce emissions consistent with observations \citep{calvet1998}. However, it still remains unclear if the magnetospheric accretion is indeed operating in real astrophysical systems. Our simulation finds that the funnel-wall accretion can be another solution that accounts for the fast, high-latitude accretion, particularly for weakly magnetized stars. Here we compare our results with previous studies and discuss implications for observations.

\subsection{Comparison with previous studies}
Historically, most of the previous studies have applied the magnetospheric accretion model to accreting stars that show fast accretion signatures. MHD modeling of accreting stars is also in accordance with this direction \citep[e.g.][]{romanova2009a,zanni2013,takahashi2017a}, although there are some studies of weakly magnetized or nonmagnetized stars \citep[e.g.][]{romanova2012,takahashi2018}. The magnetospheric accretion seems successful for explaining many observational signatures of strongly magnetized stars. However, it is not clear if the same scenario is applicable for weakly magnetized stars such as HAeBes \citep{wade2007}. Our model in which the central star is weakly magnetized suggests that a very different accretion process occurs in such a star. Nevertheless, our model shows that the funnel-wall accretion can produce observational signatures similar to those of the magnetospheric accretion. Our simulation could provide clues for solving the problem about the mechanism of fast accretion in weakly magnetized HAeBes \citep{cauley2014,reiter2017}.

We compare our results with the magnetospheric accretion model in the case of young stellar objects. We point out that a strong stellar magnetic field is not necessary to produce a funnel-wall accretion, while the strong stellar field is the key assumption in the magnetospheric accretion model. In both cases, hot spots (accretion shocks) will be formed as a result of supersonic accretion at high latitudes. The magnetospheric accretion model assumes that the accretion is guided by a magnetic field of the stellar magnetosphere, which is mainly poloidal. The accretion velocity is close to the escape velocity if the magnetosphere is sufficiently large. In our case, however, accretion streams are not guided by a stellar magnetic field, and a magnetic field is highly toroidal around the star. The maximum accretion velocity in our case is similar to the Keplerian velocity and smaller than the escape velocity by a few 10~\%, which is almost consistent with the fact that HAeBes tend to show accretion velocities systematically smaller than the escape velocity \citep{cauley2014}. The maximum latitude of accretion (and hot spots) depends on which stellar field lines guide the accreting material in the magnetospheric accretion model. However, in our case, the maximum latitude is determined by the pressure balance between the gas pressure on the disk side and the magnetic pressure around the poles. The funnel-wall accretion produces a weak wind from the region close to the star as the back reaction of the angular momentum transport, while in the magnetospheric accretion model the characteristics and existence of the wind from the region close to the star significantly depend on the magnetic field geometry and spin of the star \citep[e.g.][]{romanova2009a}.

The mass flux of the funnel-wall accretion is much smaller than that of the disk accretion. Therefore, if we only measure the mass accretion rate of the fast flows, as commonly done in the scheme of the magnetospheric accretion, we will miss most of the mass accreting onto the star. In such a case, we need to estimate the mass carried through the boundary layer as well.

The occultation of a star is observed in many accreting stars \citep{bouvier1999,cody2014,stauffer2015}. In the scheme of the magnetospheric accretion, the interpretation of the occultation is that a disk warped by the magnetosphere obscures the star \citep[e.g.][]{OSullivan2005}. We propose another possibility: the occultation by the disk surface fluctuating due to the MRI and dynamo processes. As we will see later, the density near the disk surface highly fluctuates because of the eruption of a magnetic field from the disk. This scenario may be applicable for accreting stars that show brightness variabilities not clearly related to the stellar rotation period. We again note that the accretion behavior in our simulation is highly time-variable because of the MRI-driven wind and disk dynamo.

We found the transition of the failed MRI-driven disk wind to the funnel-wall accretion, which could not be studied in previous models with a limited polar domain \citep{flock2011,suzuki2014} and local disk models \citep{bai2013a,fromang2013}. In addition, we pointed out that a continuous supply of a magnetic field from the disk is crucial for producing the strong magnetic torque around the magnetic funnels.

We comment on influences of the initial condition. Previous simulations in which a disk or torus is initially threaded by a uniform vertical magnetic field tends to produce nearly free-fall accretion near the disk/torus surface due to strong magnetic braking \citep[e.g.][]{matsumoto1996}. In our model with an hourglass-shape magnetic field, we do not find such a free-fall accretion near the disk surface. The existence of the fast disk surface accretion therefore depends on the initial condition. Unlike many MRI disk simulations, MRI-driven winds are not found in \citet{romanova2012}. Their atmospheric model based on \citet{romanova2002} assumes that the initial gas is barotropic. As a result, the gas pressure does not decrease in the vertical direction as rapid as in our model. We speculate that the high pressure above the disk prevents MRI-driven winds from blowing.


The disk extends to the stellar surface in our model, which implies the formation of the boundary layer in reality. The angular momentum transport process in the boundary layer will be different from that in the accretion disk \citep[e.g.][]{belyaev2012}. However, our damping layer approach does not allow us to study the detailed physics in that region. The boundary layer will be heated up by the dissipation of the rotational energy and can considerably contribute to the accretion luminosity. For a complete understanding of the accretion process around the star, we will improve our model to include the boundary layer in the future.

In order for a WD with the typical parameters listed in Table~\ref{tab:normalization} to establish a magnetosphere, the stellar magnetic field strength needs to be larger than several 10~kG (we estimated this from the energy balance $\rho v_{\rm K}^2/2 \approx B^2/8\pi$ near the star). However, observations show that more than 70~\% of the WDs have magnetic fields with a field strength below a few 10~kG \citep{aznarcuadrado2004,valyavin2006,scaringi2017}, which suggests that our weak stellar field model can be possibly applied to the majority of the accreting WDs. For instance, dwarf novae, binary star systems in which a WD is fed by a companion star, often show the hard X-ray emission with an energy of $\gtrsim 10$~keV at a low accretion rate ($\lesssim 10^{-10}~M_\odot$~yr$^{-1}$). Although the origin of the hard X-rays is generally attributed to the optically thin boundary layer \citep{patterson1985,narayan1993}, our model proposes an additional possible contribution to the hard X-ray emission: the emission from the accretion shocks formed by the funnel-wall accretion. Further investigations will enable us to estimate its contribution in a more quantitative way.


\subsection{Difficulty of Formation of a Magnetically Driven Jet}
A magnetically driven jet from the central region is not formed in our simulation. This result is different from the argument by \citet{kudoh1995,kudoh1997} that a magnetically driven jet with a velocity comparable to the Keplerian velocity at its foot point can be formed even if an initial poloidal magnetic field is very weak. They claimed that the generation of a magnetically driven jet is possible because the rotating gas will twist up and amplify the weak magnetic field until the magnetic energy is comparable to the gravitational energy (i.e. $v_{\it A}\sim v_{\rm K}$). However, our 3D simulation shows that a magnetic field can be amplified only before the plasma $\beta$ gets close to unity, which means that the magnetic energy is at most similar to the internal energy (or $v_{\it A} \lesssim c_{\rm s}$). The Parker instability, which can occur only in 3D, interrupts further amplification. Since the internal energy is much smaller than the gravitational energy in our cold disk (i.e. $c_{\rm s}\ll v_{\rm K}$), the magnetic energy around the field amplification regions in the disk ($z\sim 2H_{\rm p}$) is also much smaller than the gravitational energy; that is, $v_{\it A} \ll v_{\rm K}$. Therefore, the three-dimensionality and the disk thickness (or temperature) are important for the generation of a magnetically driven jet. The roles of the stellar rotation and the stellar magnetic field should be considered in future studies as well \citep[e.g.][]{hayashi1996,goodson1999a,romanova2009a}. The effects of the boundary condition and the stellar wind should also be investigated in more detail (we confirmed that the process described here does not change by performing simulations without the stellar wind).

\begin{figure}
\epsscale{.40}
\plotone{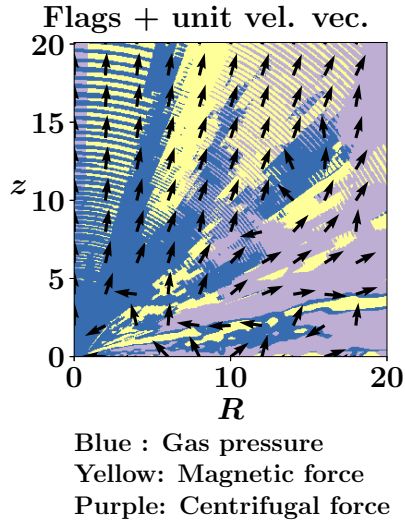}
\caption{Comparison of the magnitudes of the three forces in the direction of the poloidal magnetic fields. The gas pressure gradient, the magnetic force, and the centrifugal force are the most dominant in the blue, yellow, and purple regions, respectively. Arrows indicate the direction of the poloidal velocity only, and their length does not reflect the absolute value.}\label{fig:force_flag}
\end{figure}

To numerically check the above discussion, we examine which force is dominant for the plasma acceleration along the poloidal magnetic field, the gas pressure gradient $F_{\rm p}$ (blue), the magnetic force $F_{\rm m}$ (yellow), or the centrifugal force $F_{\rm c}$ (purple) at each location in Figure~\ref{fig:force_flag}. We define these forces as
\begin{align}
F_{\rm p} & = -\nabla p \cdot \frac{\bm{B}_{\rm pol}}{|\bm{B}_{\rm pol}|},\\
F_{\rm m} & = -\frac{1}{8\pi R^2} \frac{\bm{B}_{\rm pol}}{|\bm{B}_{\rm pol}|}\cdot \nabla(RB_{\phi})^2,\\
F_{\rm c} & = \frac{\rho v_{\phi}^2}{R}\frac{B_{ R}}{|\bm{B}_{\rm pol}|},
\end{align}
where $\bm{B}_{\rm pol}$ is the poloidal magnetic field vector \citep{tomisaka2002}. Although we can find the contribution of the magnetic force near the disk surface, the gas pressure is the main driver in a large domain within $r\lesssim10R_*$ (note that MRI has not developed in the disk of $r\gtrsim 10R_*$ until $t=300 t_{\rm K0}$). This is because the gas pressure is comparable to or larger than the magnetic force above the disk as a result of the mass loading by the MRI-driven wind and the heating by the turbulent dissipation. The magnetic-force-dominated regions appear outside.

One may expect that buoyant magnetic flux bundles continuously accelerate plasma to form a jet with a velocity comparable to the Keplerian velocity at its foot point eventually. However, the low-$\beta$ flux bundles break up and mix with the surrounding higher-$\beta$ plasma during the ascending motion, which reduces the magnetic acceleration. We infer that a disk magnetic field that is much stronger than that assumed in our model is necessary for the formation of a jet.

\subsection{Implications for protoplanetary disk evolution}
How much of the stellar radiation can reach the protoplanetary disk is crucial for the evolution of the protoplanetary disks, since the stellar radiation has an impact on the disk accretion and dispersal processes by heating materials as well as changing the ionization degree and the resistivity \citep{sano2000,hollenbach1994}. 
\citet{hirose2011} investigated radiative heating and cooling in detail in a local accretion disk using radiation MHD simulations. A detailed radiative transfer model based on a magnetospheric accretion model is given by \citet{kurosawa2013}. There are theoretical discussions that the time variability in the infrared band can be caused by materials lifted from the disk \citep{miyake2016,khaibrakhmanov2017}. 
However, because of the lack of the knowledge of the complex gas structure around the star, the complexity has been neglected in previous studies on the disk evolution. We investigate the shadowing effect of the stellar radiation due to the fluttering disk material. 

\begin{figure}
\epsscale{1.0}
\plotone{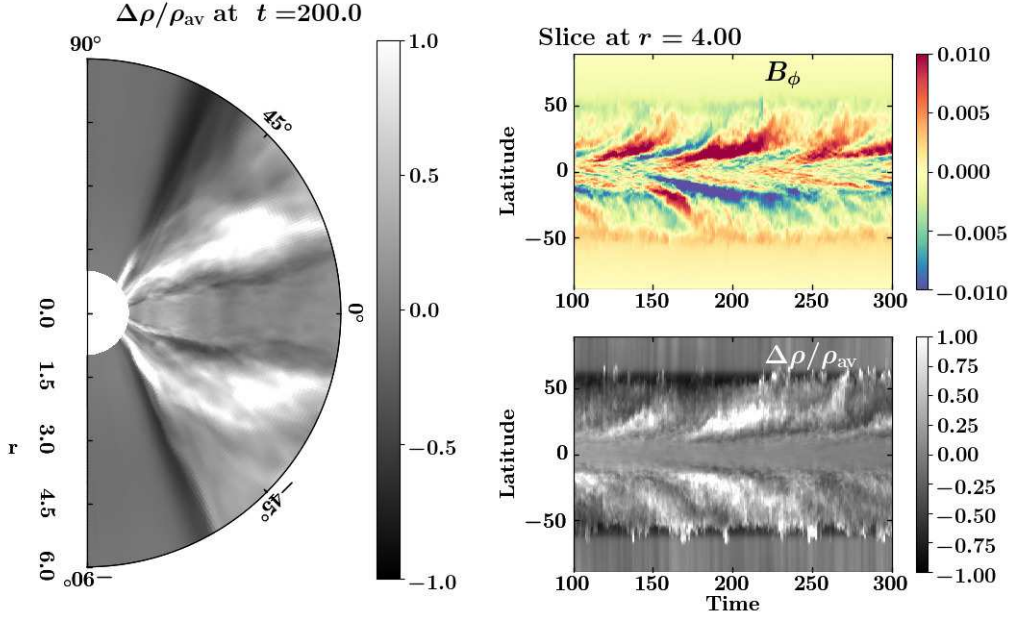}
\caption{Density fluctuation (left). Top right: time-latitude diagram of $B_{\phi}$. Bottom right: time-latitude diagram of the density fluctuation. The unit of time is $t_{\rm K0}$. An animation of this figure is available. The sequence starts at time 149$t_{\rm K0}$ and ends at time 300$t_{\rm K0}$. The animation duration is 12~s.}\label{fig:dynamo-density-fluctuation}
\end{figure}

Figure~\ref{fig:dynamo-density-fluctuation} displays the density fluctuation level around the star, defined as 
\begin{align}
\Delta \rho(r,\theta,t)/\rho_{\rm av}(r,\theta)=\left[ \rho(r,\theta,t)-\rho_{\rm av}(r,\theta)\right]/\rho_{\rm av}(r,\theta),
\end{align}
where $\rho(r,\theta,t)$ is the azimuthally averaged density, and $\rho_{\rm av}(r,\theta)$ is the temporally and azimuthally averaged density. It is evident that the density is highly fluctuating in the upper disk atmosphere. 

Figure~\ref{fig:dynamo-density-fluctuation} shows the relation between the density fluctuation in the upper disk atmosphere and the dynamo activity. The buoyant magnetic fields have a smaller density than their surroundings, and indeed we can find some examples for this: some trails of the escaping magnetic fields trace regions of the density reduction in the upper disk atmosphere (e.g. a dark trail seen between $t=150t_{\rm K0}$ and $200t_{\rm K0}$ in the lower hemisphere in the bottom right panel).

However, we also find a correspondence between some escaping magnetic fields and the enhancement of the density. 
Various processes seem to be involved with this correspondence. 
For instance, dense gas is possibly locally lifted up by buoyantly rising magnetic loops with dips (some dips are seen in Figure~\ref{fig:flux-bundle-ejection}), where materials can accumulate. The MRI-driven wind also plays a role. The MRI-driven wind becomes vigorous when strong magnetic fields appear near the disk surface as a result of the emergence or the amplification. This mass supply leads to the enhancement of the density. Note that this gas launching is highly intermittent and is driven by a combination of the magnetic pressure gradient and the magnetic tension (see Figures~\ref{fig:vertical-distribution-Eflux} and \ref{fig:slice_drho_vz}), unlike a gradual lift by rising magnetic fields subject to the Parker instability. We also notice that the radial transport (both inward and outward) of gas in the disk atmosphere can cause the density increase. Since the dynamo period depends on the radius, the timing of the Maxwell stress enhancement also changes with radius. Although the density pattern seems related to the dynamo at a local radius, this global effect makes the density pattern complex. More detailed analyses will be given in future papers.

We investigate the shadowing effect of the stellar radiation due to the fluttering disk material by estimating the column density fluctuation. Since the density fluctuation can reach $>$50\%, we can approximate the density fluctuation $\Delta n$ as $\Delta n \sim n$. The large density fluctuation occurs around the disk surface. The density in this region is typically $\sim 10^{-3}\mbox{-}10^{-2}$ times smaller than the density on the midplane at the same radius (see also the density distribution in Figure~\ref{fig:vertical-distribution}). The length scale of the fluctuations $L$ varies with time and space, but we commonly observe the fluctuation pattern with a radial scale comparable to the disk thickness. 
The disk pressure scale height for an HAeBe with a stellar mass of $3{\rm M_{\odot}}$ is $H_{\rm p} \sim 5\times 10^{10}~{\rm cm}$ at $R=5$ if the disk temperature at this radius is $10^4$~K.
When we adopt the disk thickness $2H_{\rm p}$ as the radial length scale of the fluctuation, $L \sim 10^{11}$~cm, and $n \sim 10^{14}$~cm$^{-3}$ as a typical density around the disk midplane, we estimate the column density fluctuation $\Delta N$ as 
\begin{align}
\Delta N &= \Delta n L\\
&\sim n L \\
&\sim (10^{-3}\mbox{-}10^{-2})n_{\rm mid}L\\
&= (10^{22}\mbox{-}10^{23})~{\rm cm^{-2}}\left( \frac{n_{\rm mid}}{10^{14}~{\rm cm^{-3}}}\right)\left( \frac{L}{10^{11}~{\rm cm}}\right)
\end{align}
The screening hydrogen column density (required for the optical depth $\tau$ of unity) varies with wavelength:
\begin{align}
N_{\rm H}=
\begin{cases}
10^{22}~{\rm cm^{-2}} & {\rm (X\mbox{-}ray)}\\
\ge 10^{20}~{\rm cm^{-2}} & {\rm (EUV)}\\
\ge 10^{22}~{\rm cm^{-2}} & {\rm (FUV)}
\end{cases}
\end{align}
\citep{gorti2009,nakatani2017}. The comparison demonstrates that the density fluctuation can largely affect the $\tau=1$ surface even for the radiation with these short wavelengths. Therefore, we infer that dynamo-induced density fluctuation can occult the stellar radiation that travels particularly near the disk surface. However, we note that our simulation only studied the disk with specific density and temperature profiles without solving detailed radiation processes. This could affect the actual density structure, and therefore further investigations are necessary.

Generally a warp in the inner disk caused by the stellar magnetosphere has been applied to explain a periodic photometric variability with deep, broad flux dips \citep{bouvier1999,OSullivan2005}. However, a significant fraction of the accreting young stars show aperiodic dips, which are difficult to account for with the warped disk model \citep[e.g.][]{stauffer2014,stauffer2015}. One possible explanation for this will be dust elevation due to the MRI turbulence \citep{turner2010,mcginnis2015}. We propose that the Parker instability is another possible mechanism to cause such occultation. The Parker instability occurs quasi-periodically following the dynamo period. However, the dynamo period varies with radius, and velocity and density above the disk are highly intermittent spatially and temporally due to the MRI turbulence (Figure~\ref{fig:slice_drho_vz}), which will lead to a stochastic occultation of the star.

\section{Summary}\label{sec:summary}
We presented the results of a global 3D MHD simulation of an accretion disk with a rotating, weakly magnetized star for understanding the structure of the accretion flows from an MRI-turbulent disk onto the star. We summarize our findings below:
\begin{enumerate}
\item The simulation revealed that fast accretion onto the star at high latitudes occurs even without a stellar magnetosphere. We found that the failed MRI-driven disk wind becomes the fast, high-latitude accretion as a result of angular momentum exchange mediated by magnetic fields well above the disk. Since the fast accretion occurs around the magnetic funnels, we call it the funnel-wall accretion. The funnel-wall accretion can be a solution that accounts for the fast, high-latitude accretion, particularly for weakly magnetized stars to which the magnetospheric accretion model is not applicable.
\item The funnel-wall accretion and the disk accretion around the midplane coexist. Although the speed of the funnel-wall accretion is much larger than the speed of the disk accretion, the mass accretion rate of the funnel-wall accretion is much smaller than that of the disk accretion. Therefore, if we only measure the mass accretion rate of the fast flows as commonly done in the scheme of the magnetospheric accretion, we will miss most of the mass accreting onto the star.
\item Various kinds of physics in and above the accretion disk are involved in the formation of the funnel-wall accretion. Disk materials are lifted up with the MRI-driven wind and move inward because of efficient angular momentum exchange above the disk. When the materials arrive around the magnetic funnels where magnetic fields are supplied from the inner disk through the Parker instability, the materials experience a strong decelerating torque from the Lorentz force. Since this rapid angular momentum loss occurs at a place distant from the star, their infall velocity becomes comparable to the escape velocity at the stellar surface.
\item A fast (i.e. comparable to the Keplerian velocity at its foot point), magnetically driven jet is not formed from the cold, weakly magnetized disk in our model ($H_{\rm p}/R\sim 0.1$, $\beta=10^4$). The weak magnetic field around the star is amplified by the differential rotation, but the magnetic energy cannot become comparable to the gravitational energy because the Parker instability, which only occurs in 3D disks, interrupts amplification. As a result, the magnetic fields cannot drive a jet with the velocity comparable to the Keplerian velocity at its foot point.
\item The density near the disk surface significantly fluctuates not only due to the MRI-driven wind but also due to the eruptions of the magnetic field amplified in the disk (the Parker instability). We estimated the influence of the density fluctuation on the occultation of the star, and we found that the dynamo-induced density fluctuation can largely affect the $\tau=1$ surface even for the radiation with short wavelengths (X-ray, extreme ultraviolet, and far ultraviolet). This density fluctuation process may be operating in stars that show stochastic dimming events.
\end{enumerate}

\acknowledgments
We thank Drs. K. Shibata, S. Inutsuka, H. Kobayashi, T. Hosokawa, S. Okuzumi, T. Muto, M. Kunitomo, Z. Zhu, and J. Stone for fruitful discussion. We also thank the referee for useful comments. S.T. acknowledges support by the Research Fellowship of the Japan Society for the Promotion of Science (JSPS). This work was supported in part by the Ministry of Education, Culture, Sports, Science and Technology (MEXT), Grants-in-Aid for Scientific Research, 17H01105 (T.K.S.), 16H05998 (K.T. and K.I.), 16K13786 (K.T.) and JSPS KAKENHI Grant No. 16J02063 (S.T.). Numerical computations were carried out on Cray XC30 at Center for Computational Astrophysics, National Astronomical Observatory of Japan. Test calculations were carried out on XC40 at Yukawa Institute for Theoretical Physics in Kyoto University. This research was also supported by MEXT as ``Exploratory Challenge on Post-K computer" (Elucidation of the Birth of Exoplanets [Second Earth] and the Environmental Variations of Planets in the Solar System).


\bibliography{stardisk_takasao}

\appendix

\section{Initial Atmospheric Structure}\label{apsec:ic}
The initial atmosphere is given by a self-similar solution of the axisymmetric hydrostatic equations in spherical coordinates. The density $\rho$ and the temperature $T$ are given in the forms of $\rho=r^{-a}f(\theta)$ and $T=P/\rho=r^{-1}g(\theta)$, respectively. The hydrostatic balances in the $r$ and $\theta$ directions lead to 
\begin{align}
\frac{\partial p}{\partial r} - \frac{\rho v_{\rm \phi}^2}{r} & = -\frac{\rho}{r^2}, \label{eq:r-balance}\\
\frac{\partial p}{\partial \theta} &= \rho v_{\rm \phi}^2 \frac{\cos{\theta}}{\sin{\theta}},\label{eq:theta-balance}
\end{align}
respectively, where the gravitational constant $ G$ is taken so that $GM_{*}=1$.
We obtain the rotational velocity distribution $v_\phi$ from Equation~\ref{eq:r-balance} as follows:
\begin{align}
v_\phi &= \sqrt{-(a+1)T + \frac{1}{r}}\\
&=\sqrt{\frac{1-(a+1)g(\theta)}{r}}
\end{align}
This expression does not produce any unphysical discontinuity of the angular velocity at the boundary between the disk and the disk atmosphere. Note that this equation gives the upper limit of the function $g(\theta)$.
The hydrostatic balance in the $\theta$ direction (Equation~\ref{eq:theta-balance}) yields
\begin{align}
\frac{d\ln{\left[f(\theta)g(\theta)\right]}}{d\theta} = \frac{\cos\theta}{\sin\theta}\left[ \frac{1}{g(\theta)} - (a+1)\right].  \label{eq:ic_eq}
\end{align}
We solve this equation with a given form of the function $g(\theta)$ and an appropriate boundary condition at $\theta=\pi/2$ (midplane). This equation gives $f(\theta)$, and the density distribution. The boundary condition for $f(\theta)$ is $f(\pi/2)=\rho_{\rm disk,0}$,where $\rho_{\rm disk,0}$ is the density at the disk midplane at the radius of 1, and we set $\rho_{\rm disk,0}=10$. We adopt $a=2$ for the power-law index of the density.

We determine the form of $g(\theta)$ so that the initial atmosphere consists of the cold disk and the hot atmosphere above it. The functional form is given by
\begin{align}
g(\theta) = T_{\rm disk,0}\left\{  \frac{k_{\rm c}-1}{2}\left[ \tanh\left( \frac{\theta-\theta_2}{\Delta \theta_{\rm TR}}\right) - \tanh\left( \frac{\theta-\theta_1}{\Delta \theta_{\rm TR}}\right) + 2\right] + 1 \right\},
\end{align}
where $\theta_{1,2}$ are the longitudinal angles of the upper and lower transition region between the disk and the corona, respectively. $\Delta \theta_{\rm TR}$ denotes the typical range of the angle of the transition region. Here, we assume that $T(r,\pi/2)=T_{\rm disk,0} r^{-1}$, where $T_{\rm disk,0}$ is the temperature at the disk midplane at the radius of 1. Note that $k_{\rm c}$ is the factor representing the jump in the temperature between the disk and corona and should be smaller than $1/T_{\rm disk,0}(a+1)$. We set $k_{\rm c}=0.9/T_{\rm disk,0}(a+1)$ in this study. We introduce a small random perturbation to the pressure whose amplitude is 1~\% of the local pressure. We set $\theta_{1,2}=\pi/2 \mp \Delta \theta_{\rm disk}$, respectively, where $\Delta \theta_{\rm disk}=4\tan^{-1}{(H_{\rm p}/r)}$. Here, $\Delta \theta_{\rm TR}=\tan^{-1}{(H_{\rm p}/r)}$, and $T_{\rm disk,0}=0.01$.

The boundary condition of Equation~\ref{eq:ic_eq} is given by the gas pressure at the midplane. We assume that $\rho(r,\pi/2)=\rho_{\rm disk,0} r^{-a}$. From the temperature and density distributions, the boundary condition becomes
\begin{align}
f(\pi/2)g(\pi/2) = \rho_{\rm disk,0} T_{\rm disk,0}.
\end{align}
By solving Equation~\ref{eq:ic_eq} with the above boundary condition, we can get the density distribution.

The initial poloidal magnetic field is an hourglass-shape field described in \citet{zanni2007}. The $\phi$ component of the vector potential is written as 
\begin{align}
A_{\phi}(r,\theta) = \frac{2B_{\rm z,0}}{3-a} r^{-(a-1)/2} \left[ 1+\frac{1}{m^2} \left( \frac{z}{r} \right)^2 \right]^{-5/8}.
\end{align}
The star has a weak magnetic field according to this vector potential. The density distribution with the initial field lines is shown in Figure~\ref{fig:ic}. The poloidal field is derived as $\bm{B}=\nabla \times (A_\phi \hat{\phi})$. The field strength at the midplane scales as $B_{\rm z,0}r^{-(a+1)/2}$, which gives a constant plasma $\beta$ at the disk midplane, the ratio of gas to magnetic pressure, for the adopted density and temperature distributions. The field strength $B_{\rm z,0}$ is written as $B_{\rm z,0} = \sqrt{8\pi\rho_{\rm disk,0}T_{\rm disk,0}/\beta_0}$, where $\beta_0$ is the initial plasma $\beta$ on the disk midplane. The parameter $m$ characterizes the length scale on which the magnetic field bends ($m \rightarrow \infty$ gives a perfectly vertical field). In this study, we set $\beta=10^4$ and $m=1$.

\section{Radiative Cooling in the Disk}\label{apsec:radiative-cooling}
We include a simplified radiative cooling to explore a quasi-steady state established with the initial disk temperature profile. We solve the following equation via operator splitting:
\begin{align}
\frac{\partial T}{\partial t} &=-\frac{T(r,\theta)-T_0(r,\theta)}{\tau_{\rm cool}(r,\theta)},
\end{align}
where $T_0(r,\theta)$ is the initial temperature profile. To switch on the radiative cooling only in the disk region, we define the radiative cooling timescale $\tau_{\rm cool}(r,\theta)$ as
\begin{align}
\tau_{\rm cool}(r,\theta)^{-1} & = \frac{\Omega_{\rm K}(r)}{2\pi f_{\rm cool}} F_{1}(\theta) F_{2}(\rho),\\
F_{1}(\theta)&=\frac{1}{2}\left[ 1-\tanh{\left( \frac{\theta^\prime - \Delta\theta_{\rm disk}}{\Delta\theta_{\rm cool}}\right)} \right],\\
\theta^\prime &= \left|\frac{\pi}{2} - \theta \right|, \\
F_{2}(\rho)&=\frac{1}{2}\left[ 1+\tanh{\left( \frac{\rho-\rho_{\rm cool}}{\rho_{\rm cool}}\right)} \right].
\end{align}
This functional form switches on the radiative cooling term only in the disk region ($|\pi/2-\theta| < \Delta \theta_{\rm disk}$, $\Delta \theta_{\rm cool} = 0.5\tan^{-1}{(H_{\rm p}/r)}$) with a moderate density ($\rho > \rho_{\rm cool}$). In this study, $\rho_{\rm cool}$ is set to $10^{-4}$. The radiative cooling timescale is determined by a parameter $f_{\rm cool}$. We set $f_{\rm cool}=0.2$, which means that the radiative cooling timescale is 20\% of one orbital period. We confirmed that the temperature profile near the disk midplane is maintained without a significant change due to viscous heating.

\section{Damping Layer Method for the Inner Boundary}\label{apsec:damping}
We model the stellar surface to simulate the accretion onto the central star. 
Our stellar surface model is constructed to satisfy the following requirements: (1) the stellar surface should be rigid in the sense that the falling material cannot freely penetrate into the stellar interior, (2) the accreting material will be absorbed by the star eventually but gradually, and (3) the (thermally driven) stellar wind blows from the hot stellar corona. To model this situation, neither a reflecting boundary nor an outgoing (or free) boundary is appropriate.

Although the stellar surface is disturbed by the accretion flows, the stellar surface should revert to the original state from the disturbed state. To model this situation, we construct a damping layer method described below. The basic concept of a damping layer is that the physical quantities in a defined domain are controlled to approach specified values in a spatially and temporally smooth way. Although not perfect, adopting the damping layer is an acceptable compromise because it is extremely challenging with computational resources available today to simulate the disk and outflow structures on an astronomical unit scale while at the same time resolving photospheric structures on a scale of a few hundred kilometers. Our stellar ``surface'' is located at the radius of 1, and we put a thin damping layer between this radius and the radius of the inner boundary (i.e., $0.91R_{*}<r<R_{*}$ in this study). We consider that our stellar surface represents the position of the bottom of the corona, and the thermal property of the coronal gas is mainly controlled by the lower layer, that is, the damping layer.

In the damping region, we solve the following equations in addition to the basic equations via operator splitting:
\begin{align}
\frac{\partial \rho}{\partial t} &= - \frac{\rho-\rho_{\rm star}}{\tau_{\rm d}(r,\rho)},\\
\frac{\partial \bm{v}}{\partial t} & = -\frac{\bm{v}-\bm{v}_{\rm star}}{\tau_{\rm d}(r,\rho)},\\
\frac{\partial p}{\partial t} & = -f_{\rm s}(S,\rho)\frac{p-p_{\rm star}}{\tau_{\rm d}(r,\rho)}, \label{eq:pdamp}
\end{align}
where $\rho_{\rm star}$ and $p_{\rm star}$ denote the stellar coronal density and pressure, respectively. The total energy per unit volume is recalculated using the updated quantities. We set the stellar temperature $T_{\rm star}(=p_{\rm star}/\rho_{\rm star})$ to be the virial temperature (1 in this study) so that the stellar wind blows. With the angular velocity of the rotating star $\Omega_{*}$, the stellar rotational velocity is defined as $\bm{v}_{\rm star}=(0, 0, \Omega_{*} r \sin{\theta})$ (the rotation axis of the star coincides with the rotational axis of the accretion disk). Here, $\tau_{\rm d}(r,\rho)$ is the damping timescale and a function of the spherical radius and the density. The function $f_{\rm s}(S,\rho)$ is a function of the specific entropy $S=\ln{(p/\rho^{\gamma})}$ and the density and prevents an artificial thermal convection driven by the heating in the damping layer. The definition of this function will be mentioned later.

We take the sound crossing timescale in the damping layer $t_{\rm cross}=w_{\rm d}/c_{\rm star}$ as the unit of the damping timescale, where $w_{\rm d}$ is the thickness of the damping layer ($0.09R_{*}$), and $c_{\rm star} = \sqrt{T_{\rm star}}$. The damping layer should be smoothly connected from the stellar surface. We assume that the damping timescale depends on the density, because the star will absorb higher density materials at a longer timescale. Considering these requirements, we adopt the following functional form for the damping timescale $t_{\rm d}$:
\begin{align}
t_{\rm d}(r,\rho)^{-1} &= f_{\rm rad}(r)f_{\rm v}(v_{r})t_{\rm d0}(\rho)^{-1},\\
t_{\rm d0}(\rho) &= \max\left[ \min\left(f_{\rm d,min}\frac{\rho}{\rho_{\rm star}}, f_{\rm d,max} \right), f_{\rm d,min}\right] \times t_{\rm cross} \label{eq:td0}, \\ 
f_{\rm rad}(r) & = \frac{1}{2}\left[1-\tanh{\left( \frac{r-r_{\rm d}}{w_{\rm rad}} \right)} \right],\\
f_{\rm v}(v_{ r}) & = \frac{1}{2}\left[ \tanh{\left( \frac{v_{ r}-v_{r,{\rm c}}}{0.1|v_{r,{\rm c}}|}\right)} + 1\right].
\end{align}
Here, $t_{\rm d0}$ represents the dependency of the damping timescale on the density. The value smoothly changes with the density and is limited within $f_{\rm d,min}t_{\rm cross}$ and $f_{\rm d,max}t_{\rm cross}$.
The damping layer is smoothly connected with a function $f_{\rm rad}(r)$. Here, $r_{\rm d}$ is set to $0.91R_{*}+w_{\rm d}/2$, and $w_{\rm rad}=2w_{\rm d}/15$. We also introduce a function $f_{\rm v}(v_{r})$ so that the velocity of the fast accreting materials that can produce shocks is not damped before they collide with the inner boundary to bounce back. We set the floor radial velocity $v_{r,{\rm c}} = -0.05 v_{\rm K0}$.

Equation~\ref{eq:td0} means that the damping timescale is $f_{\rm d,min}t_{\rm cross}$ in a low-density region like the stellar wind base, while $f_{\rm d,max}t_{\rm cross}$ in a high-density region like the star-disk boundary. Since the wind speed is almost the coronal sound speed, the damping time for the stellar wind base should be comparable to the sound crossing time $t_{\rm cross}$. This requires that $f_{\rm d,min}\approx 1$ or smaller. However, such a rapid damping of high-density regions (e.g. the star-disk boundary) results in some unphysical results in our experiments, such as an unphysical convection. Therefore, we choose $f_{\rm d,max}=10^3$ on the basis of our numerical experiments. It turned out that a large value of $f_{\rm d,max}$ is insufficient to suppress the unphysical convection. For this reason, we introduce the function $f_{\rm s}(S,\rho)$ in the pressure damping equation (Equation~\ref{eq:pdamp}) so that the pressure damping (which mostly acts as heating) does not operate in the regions of negative entropy gradient. The adopted functional form is
\begin{align}
f_{\rm s}(S,\rho) &= 1 + \left[ f_{\rm s0}(S) - 1 \right] f_{\rm \rho}(\rho),\\
f_{\rm s0}(S) & =
\begin{cases}
0 & (dS/dr < 0)\\
1 & ({\rm otherwise}),\\
\end{cases}\\
f_{\rho}(\rho) & = \frac{1}{2}\left[ \tanh{\left( \frac{\rho-\rho_{\rm d}}{0.1\rho_{\rm d}}\right) } + 1\right].
\end{align}
This functional form allows us to switch off the damping of pressure in the high-density and negative entropy gradient regions.
We add a function $f_{\rho}(\rho)$ to switch on the damping at the stellar wind base.
Note that $f_\rho(\rho)$ goes to 0 in the regions with the density smaller than $\rho_{\rm d}$, and to 1 in the denser regions. Here, $\rho_{\rm d}$ characterizes the coronal region, and we take $\rho_{\rm d}=3\rho_{\rm star}$. 

In this formulation, the accretion region and the stellar coronal region (stellar wind region) at the stellar surface are automatically determined. The mass accretion rate onto the star and the mass loss rate of the stellar wind are not explicitly related to each other, although they may be so in reality.

\section{Resolution Diagnostics}\label{apsec:resolution}
We examined whether the MRI turbulence is well captured in our simulation. We evaluated a quality factor $Q_i$ of the $i$ component for MRI ($i=r$, $\theta$, and $\phi$) \citep{noble2010,hawley2013}, which is defined as the ratio of the characteristic wavelength of the MRI mode to the grid size:
\begin{align}
\langle Q_{i}(r,\theta) \rangle_{\phi} = 2\pi \frac{\sqrt{\langle v_{{\it A},i}(r,\theta)^2 \rangle}}{\Omega_{\rm K} \Delta l_{i}},
\end{align}
where $\langle v_{\it A, { i}}^2\rangle_{\phi} = \langle B_{i}^2\rangle_{\phi} / 4\pi \langle \rho \rangle_{\phi}$ is the square of the Alfv\'en speed based on the $i$ component of the magnetic field, and $\Delta l_i$ is the grid size in the $i$ direction and $\Delta l_i = \Delta r$, $r\Delta \theta $, and $r\sin{\theta} \Delta \phi$ for the $i=r$, $\theta$, and $\phi$ components, respectively.

\begin{figure}
\epsscale{1.2}
\plotone{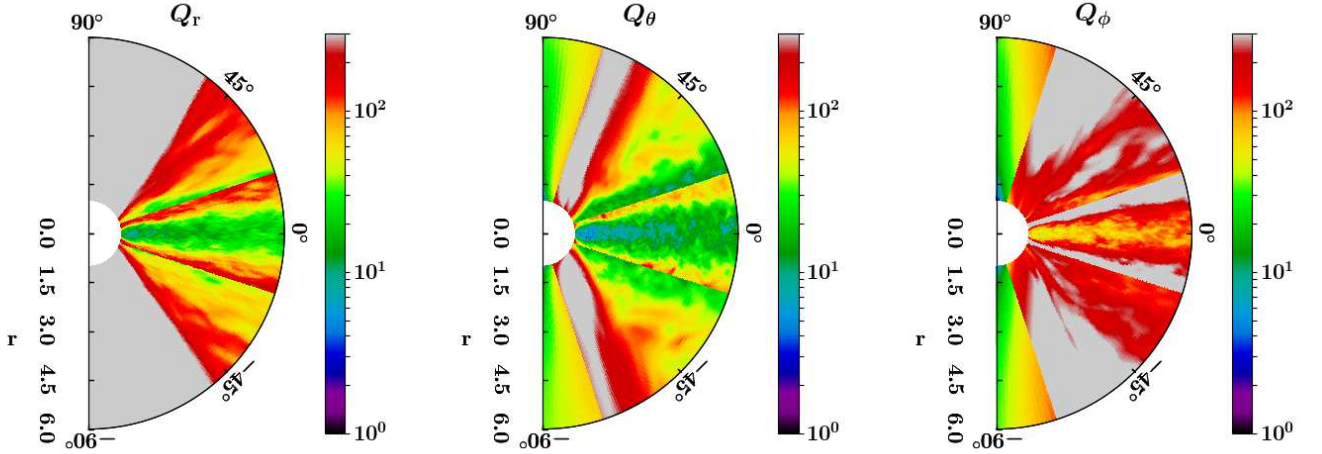}
\caption{From left to right, snapshots of the quality factors $Q_{\rm r}$, $Q_{\theta}$, and $Q_{\rm \phi}$ at $t=283t_{\rm K0}$ near the star (within $r=6$). All the quantities are averaged in the azimuthal direction.}\label{fig:q-factors}
\end{figure}

Figure~\ref{fig:q-factors} shows the three components of the quality factors. The values of $Q_{r}$ and $Q_{\phi}$ are larger than 10 in the disk, which means that the MRI modes in these two directions are well resolved. The value of $Q_\theta$, on the other hand, is around 10 and sometimes becomes several in the very inner disk. \citet{sano2004} argued that $Q_\theta \gtrsim 6$ is necessary to obtain an MRI turbulence whose characteristics are similar to that in well-resolved cases. Although our simulation almost meets this condition, the criterion $Q_\theta \ge 15 $, introduced by \citet{hawley2013} is sometimes not satisfied in the very inner disk. However, the wind launching regions (the disk surface around $\pm 16^\circ$ from the midplane) and the funnel-wall accretion regions (around $\pm 45^\circ$ from the midplane) are well resolved in terms of the quality factors. We also confirmed that the funnel-wall accretion is formed in simulations with a coarser spatial resolution. Therefore, we consider that the funnel-wall accretion is real. However, since the spatial resolution in the $\theta$ direction may not be sufficient, higher-resolution calculations will be necessary for a more quantitative discussion.

\end{document}